\documentclass[journal,10pt]{IEEEtran}
\ifCLASSINFOpdf
\else
\fi
\usepackage{cite} 
\usepackage{amsmath}
\usepackage{graphicx}
\usepackage{balance}

\usepackage{multirow}
\usepackage{color}

\usepackage{caption}
\usepackage{subcaption}
\usepackage{stfloats}
\usepackage{mathtools}

\usepackage{tcolorbox}

\begin{document}
%
\title{FlexONC: Joint Cooperative Forwarding and Network Coding with Precise Encoding Conditions}
%
%
%


\author{Somayeh Kafaie, Yuanzhu Chen, Mohamed Hossam Ahmed,
        and~Octavia A. Dobre
\thanks{S. Kafaie, M. H. Ahmed, and O. A. Dobre are with the Faculty of Engineering and Applied Science, Memorial University of Newfoundland, St. John's, NL, A1B 3X5, Canada (e-mail: somayeh.kafaie@mun.ca, mhahmed@mun.ca, odobre@mun.ca}
\thanks{Y. Chen (correspondence author) is with the Department of Computer Science, Memorial University of Newfoundland, St. John's, NL, A1B 3X5, Canada (e-mail: yzchen@mun.ca)} }

\maketitle

\begin{abstract}
In recent years, network coding has emerged as an innovative method that helps a wireless network approach its maximum capacity, by combining multiple unicasts in one broadcast. However, the majority of research conducted in this area is yet to fully utilize the broadcasting nature of wireless networks, and still assumes fixed route between the source and destination that every packet should travel through. This assumption not only limits coding opportunities, but can also cause buffer overflow in some specific intermediate nodes. Although some studies considered scattering of the flows dynamically in the network, they still face some limitations. This paper explains pros and cons of some prominent research in network coding and proposes a Flexible and Opportunistic Network Coding scheme (FlexONC) as a solution to such issues. Furthermore, this research discovers that the conditions used in previous studies to combine packets of different flows are overly optimistic and would affect the network performance adversarially. Therefore, we provide a more accurate set of rules for packet encoding. The experimental results show that FlexONC outperforms previous methods especially in networks with high bit error rate, by better utilizing redundant packets spread in the network.
\end{abstract}


%
\IEEEpeerreviewmaketitle

\section{Introduction}
%
%
%
%
\IEEEPARstart{I}{n} recent years, a significant amount of research has been conducted to explore the effect of network coding in different scenarios and improve the network performance. To exploit network coding, related research mostly focuses on either inter-flow or intra-flow network coding. 

\begin{figure}
\centering
\begin{subfigure}{.3\textwidth}
  \centering
  \includegraphics[width=0.8\linewidth]{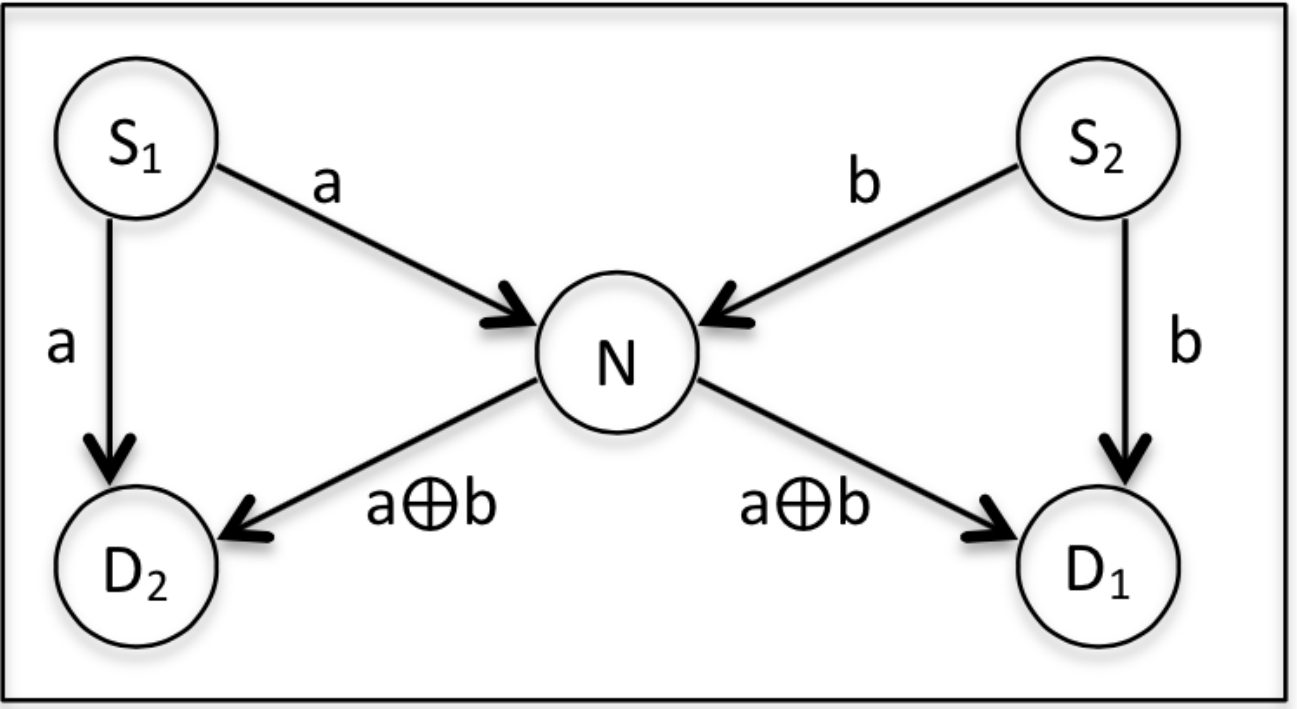}
  \caption{X-topology.}
  \label{fig:Xtopology}
\end{subfigure}%
\begin{subfigure}{.2\textwidth}
  \centering
  \includegraphics[width=0.8\linewidth]{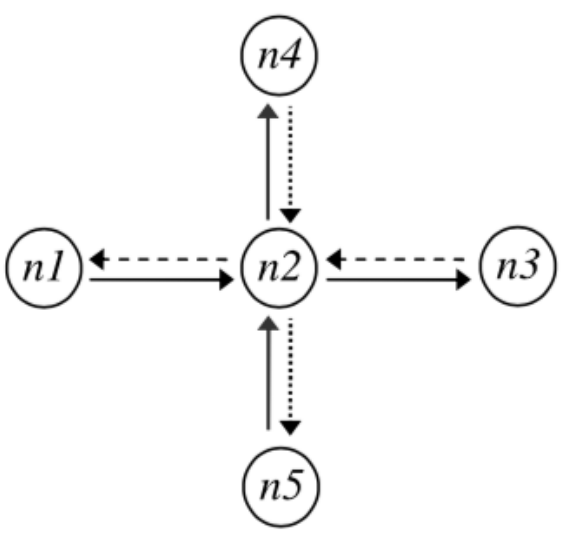}
  \caption{Cross topology~\cite{COPE-Katti-IEEEACMTransactions2008}.}
  \label{fig:cross}
\end{subfigure}
\caption{Some topologies utilizing network coding.}
\label{fig:NC-topologies}
\end{figure}

One of the most popular examples showing the gain behind inter-flow network coding is the X-topology in \figurename~\ref{fig:Xtopology}, where $S_{1}$ sends packet $a$ to $D_{1}$, and $S_{2}$ sends packet $b$ to $D_{2}$، through an intermediate node $N$. Since $D_1$ and $D_2$ are able to overhear the packets of the other flow from its source, the relay node $N$ mixes packets of two flows and sends their combination to the network. Doing so, network coding decreases the number of required transmissions to deliver packets to their final destination and improves the performance.

COPE~\cite{COPE-Katti-IEEEACMTransactions2008} is one of the first methods that realize this idea in practical scenarios. Whenever an intermediate node receives packets from different flows, it encodes them if it is likely that the next-hops of the native packets combined in the coded packet are able to decode this packet and retrieve the original content. However, coding opportunities in COPE are restricted only to joint nodes that receive packets from multiple flows. Therefore, to provide more coding opportunities, COPE needs more packets to arrive at the same node. However this traffic concentration may overload intermediate nodes, and cause longer delay, buffer overflow, and channel contention.

As a solution to this problem, BEND~\cite{BEND-Zhang-CNJournal2010} applies network coding while trying to avoid traffic concentration. By taking advantage of the broadcasting nature of wireless networks, BEND allows all receivers of the packet, in addition to the intended next-hop specified by the routing protocol, to help in mixing and forwarding the packet if they believe they can be helpful. However, these \emph{non-intended forwarders} (i.e., the receivers of the packet which are not specified as the next-hop on the route defined by the routing protocol, and can help in forwarding) are allowed to assist the \emph{intended forwarder} only in forwarding received native packets. In fact, if they receive a coded packet, they just discard it, even if they were able to decode the received packet. This restriction not only limits the number of coding opportunities in the network but also increases the number of retransmissions. The terms \emph{intended} and \emph{non-intended forwarders} as well as some other terms used in this research are summarized in Table~\ref{table:definition}.

Furthermore, almost all inter-flow network coding methods, which mix packets within a two-hop region, follow a similar set of coding conditions to encode packets. We call this set ``common coding conditions''. Based on these coding conditions, given a high delivery probability between nodes, two packets are combined if the next-hop of each packet is the previous hop of the other packet or one of the neighbors of the previous hop. However, in some scenarios as shown in this research, the common coding conditions may decide incorrectly to mix some packets that cannot be decoded at the next-hops. This wrong encoding causes failures in decoding, increases the number of required retransmissions to deliver the packets and consequently decreases the network throughput.

To better utilize the broadcasting nature of wireless networks, we introduce FlexONC (Flexible and Opportunistic Network Coding), which provides more flexibility to previous methods like COPE and BEND by adding opportunistic forwarding, and allowing non-intended forwarders to help in decoding in addition to encoding and forwarding. Moreover, FlexONC proposes an additional coding condition to find coding opportunities more accurately, and designs a mechanism to merge it with the common coding conditions.

The main contributions of FlexONC are as follows: 1) More diffusion gain since more packets (i.e., coded and native packets) can be forwarded by a node other than their intended forwarder; 2) Faster packet delivery to the final destination because even if the intended forwarder does not receive the packet or cannot decode the received coded packet, some non-intended forwarders can still help; 3) More coding opportunities as non-intended forwarders are eligible to receive and probably decode coded packets and consider them as candidates to be mixed with other packets; 4) More intelligent and comprehensive encoding decisions to avoid transmitting undecodable packets in the network.

The rest of the paper is organized as follows. Related research on network coding, especially COPE and BEND, is discussed in Section~\ref{section:relatedWork}. Section~\ref{section:examples} provides two examples to show the effectiveness of FlexONC. Section~\ref{section:design} describes the objectives and challenges of FlexONC, and introduces its implementation details. Section~\ref{section:performance} presents performance evaluation results and compares FlexONC with a non-coding scheme as well as other inter-flow network coding methods. In Section~\ref{section:discussion}, some intrinsic features of FlexONC are discussed further. Finally, Section~\ref{section:conclusion} concludes the paper and provides ideas to extend FlexONC in future research.

\begin{table}[!t]
\renewcommand{\arraystretch}{1.3}
\caption{Definition of some terms used in this article.}
\label{table:definition}
\centering
\begin{tabular}{|l||l|}
\hline
\textbf{Term} & \textbf{Definition} \\ \hline
native packet &  a packet that is not combined with any other packet \\  \hline
coded packet &  XORed of more than one native packet \\  \hline
intended forwarder & the designated next-hop by the routing protocol \\  \hline
non-intended &  the neighbors of the next-hop \\
forwarder & which can help in forwarding \\  \hline
coding node &  a node in which coded packets are generated \\  \hline
eligible & a node which is the neighbor of both the \\
forwarder & next-hop and the second next-hop of a packet \\  \hline
decoded-native &  a native packet which was received coded \\
packet &  and has been decoded \\  \hline
coding partner &  each native packet encoded with other packets \\ \hline
 common coding &  the conditions used by previous methods  \\
  conditions & (e.g., COPE and BEND) to combine packets \\ \hline
\end{tabular}
\end{table}

\section{Background and Related Work} \label{section:relatedWork}

Network coding represents an innovative idea introduced by Ahlswede \textit{et al.}~\cite{NC-Ahlswede-IEEETransactionsIT2000} in 2000 to increase the transmission capacity of the network, as well as its robustness. In general, two different types of network coding can be applied, namely intra-flow and inter-flow network coding. While in the former, nodes mix packets of the same flow to increase the robustness~\cite{MORE-Chachulski-SIGCOMM2007, E-NCP-Lin-INFOCOM2008, ICEMAN-Joy-MobiCom2013}, in the latter packets of different flows are mixed to reach the maximum capacity of the network~\cite{COPE-Katti-IEEEACMTransactions2008, BEND-Zhang-CNJournal2010, NC+DTN-Widmer-SIGCOMM2005}. Xie \textit{et al.} provide a survey on inter-flow network coding under both reliable links and lossy links~\cite{XNCSurvey-Xie-ComNet2015}.

In inter-flow network coding, an intermediate node combines two packets if the next-hop of each packet has already received the other coding partner. To keep track of the packets received by each node, two types of information are used: deterministic information and probabilistic information. Deterministic information are provided by exchanging ``reception reports'' among nodes, where each node's reception report contains the packets that recently have been received or overheard by the node~\cite{COPE-Katti-IEEEACMTransactions2008}. These reception reports are usually piggybacked on data packets or broadcasted periodically. 

In the absence of deterministic information (e.g., when a node does not transmit any data packet and only relies on periodic updates), probabilistic information is used to decide on encoding. In this case, if the delivery probability between nodes is greater than a threshold, two packets are combined if the next-hop of each packet is the previous-hop of the other coding partner or one of the neighbors of the previous-hop. In this research, we present scenarios where encoding decisions made based on the probabilistic information through the common coding conditions are not accurate enough and cause a significant number of decoding failures.  

COPE is one of the prominent examples of inter-flow network coding. In COPE, a node combines the packets, $P_{1}$, $P_{2}$, ..., $P_{n}$, with different next-hops, $NH_{1}$, $NH_{2}$, ..., $NH_{n}$, when in the combined packet 1) for each next-hop there is at most one packet, and 2) for each packet $P_i$, all the next-hops have already received the packet except for its corresponding next-hop, $NH_i$. For example, let us assume that in the cross topology depicted in \figurename~\ref{fig:cross}, for each node all nodes are in its transmission range except for the diametrically opposed node, and $n_1$, $n_3$, $n_4$ and $n_5$ are the sources of 4 flows intersecting at $n_2$. Then, $n2$ can mix 4 packets received from all sources because each next-hop contains all other coding partners except for its intended packet. However, the improvement of throughput in COPE depends on the traffic pattern. In fact, it limits coding opportunities because coding can be accomplished only at joint nodes. As an example, if in \figurename~\ref{fig:cross} the sources choose a different intermediate node than $n_2$, all flows cannot intersect at the same node and less coding opportunities are provided by COPE.  


A variety of improvements over COPE have been put forward, especially by adding opportunistic forwarding~\cite{NCSurvey-Iqbal-NCAJournal2011}. In CORMEN~\cite{CORMEN-Islam-COMPUTING2010}, as a network coding scheme enhanced with opportunistic routing, the nodes in the forwarder set are neighbors of the nodes in the shortest path to avoid diverging the path and unnecessary duplicate packets. However, similar to source routing protocols, the packet header should contain not only the forwarder set but also the nodes in the shortest path. In addition, since the packet may not follow the shortest path, the forwarders need to keep updating it. Also, end-to-end acknowledgments are sent instead of hop-by-hop ones.

CORE~\cite{CORE-OR-Yan-IEEEWC2010} is also one of the first reaserch that integrates inter-flow network coding with opportunistic forwarding to increase the coding opportunities in the network. In each transmission, among all neighbors of the last forwarder which are closer than it to the destination, CORE selects the node with more coding gain as the next forwarder. To prioritize the nodes with different coding opportunities, forwarding timers are used so that the node with more coding opportunities forwards its packet earlier. In addition, in CORE the packets are broadcated without any acknowledgment and retransmission mechanism. While CORE defines a coding gain function in each node only in terms of the number of neighbors that are able to decode a coded packet, CoAOR~\cite{CoAOR-Hu-GLOBECOM2013} takes into account the number of flows coded in a packet and the link quality as well. 

CAR~\cite{CAR-Liu-NetSysManage2015} is another coding-aware opportunistic routing scheme that aims to maximize the number of native packets coded together in a single transmission by dynamically selecting the route based on real-time coding opportunities. In some described works, the closeness to the destination (i.e., to find the forwarding set) is calculated in terms of the geographical distance, which does not necessarily represent the quality of the path. In addition, in most of the research in this area, the maximum coding opportunities is the only factor taken into account to select the next forwarder, even if the path traveled by the node is excessively longer than the shortest path.

BEND, as another advancement of COPE, introduces a type of gain, referred to as the \emph{diffusion gain}, which is the benefit of being able to scatter flows through multiple forwarders dynamically. In BEND, each node has three queues: $Q_{1}$ for intended native packets, $Q_{2}$ for overheard native packets, and \emph{mixing-Q} for coded packets. A node can combine two packets if the next-hop of the first packet is the previous hop of the second packet or one of its neighbors, and vice versa.

\begin{figure}[ht]
\centering
\includegraphics[scale=0.5]{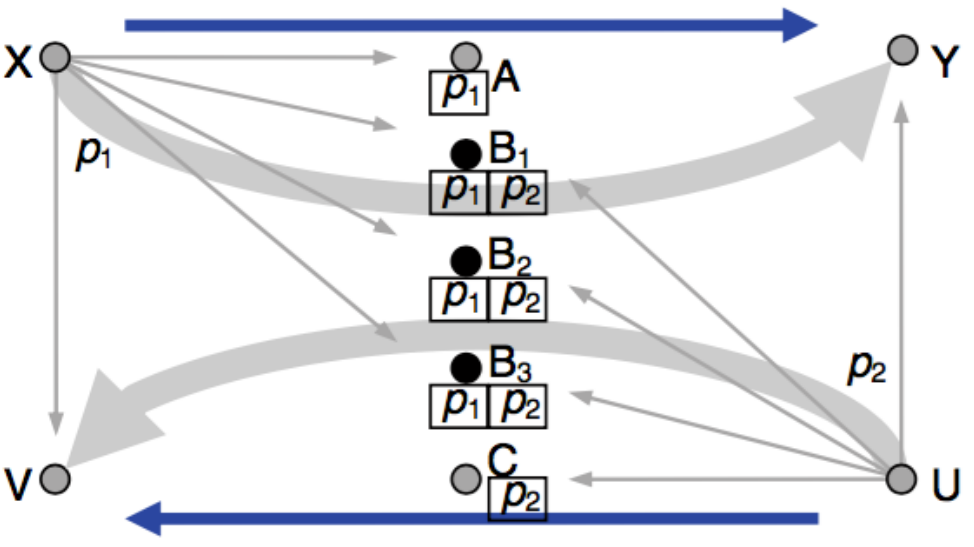}
\caption{Diffusion gain in BEND~\cite{BEND-Zhang-CNJournal2010}.}
\label{figure:diffusion}
\end{figure}

To avoid traffic concentration in BEND, a non-intended forwarder may receive a native packet and mix and forward it on behalf of the intended forwarder. For example in \figurename~\ref{figure:diffusion}, where $A$ and $C$ are the intended forwarders of the flows from $X$ and $U$ to $Y$ and $V$, respectively, COPE cannot find any coding opportunity. On the other hand, BEND allows non-intended forwarders which can overhear packets of both flows (e.g., $B_1$, $B_2$ and $B_3$) to combine and forward the packets on behalf of the intended forwarders. To do so, a \emph{second-next-hop} field is included in native packets. As such, when a non-intended forwarder receives a native packet, it can find the address of the next-hop in the \emph{second-next-hop} field. 

\begin{figure}[ht]
\centering
\includegraphics[scale=0.4]{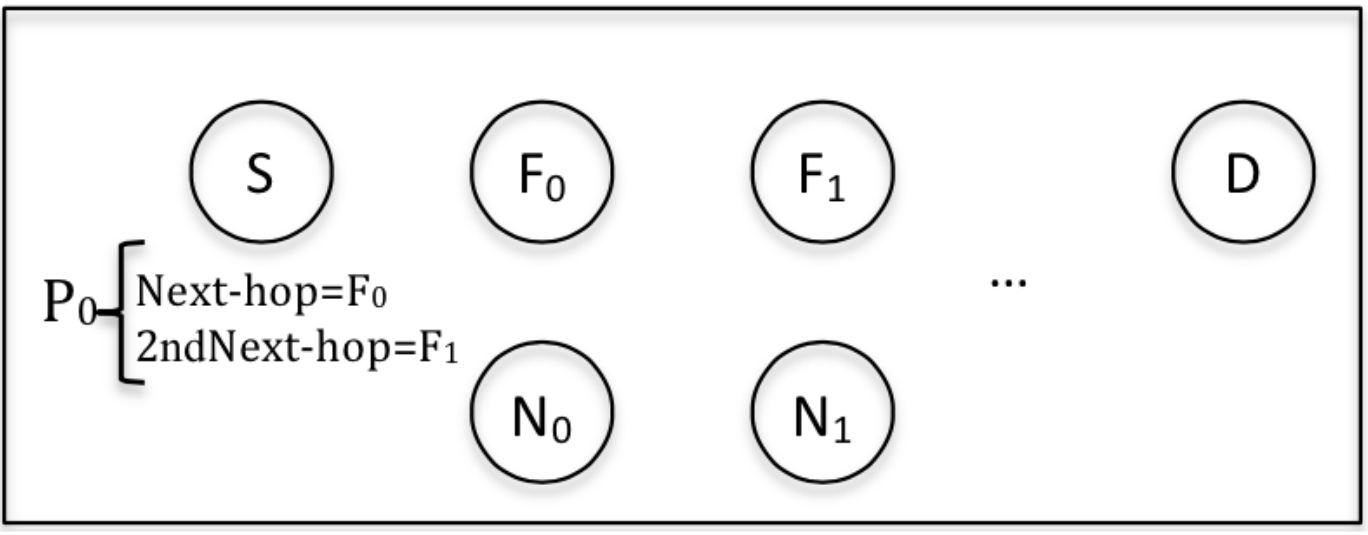}
\caption{In BEND, non-intended forwarders drop coded packets.}
\label{figure:SNH}
\end{figure}

However for coded packets, the \emph{second-next-hop} field does not present the correct address in a way that the packets still travel near the original route. Therefore, non-intended forwarders must drop coded packets since they do not know the address of the next-hop from the intended forwarder to the destination. To illustrate the idea, let us assume in \figurename~\ref{figure:SNH} that the source $S$ sends a packet $P_{0}$ to $D$. Based on the information provided by the routing protocol, it fills the \emph{next-hop} and \emph{second-next-hop} fields with $F_{0}$ and $F_{1}$, respectively. We assume that $F_{0}$ fails to receive the packet, and $N_{0}$ overhears it. In addition, $N_{0}$ can mix $P_{0}$ with a packet $P_{1}$ in its buffer and forward it. Based on $P_{0}$'s header, $N_{0}$ sets the new \emph{next-hop} field with the current \emph{second-next- hop} field, $F_{1}$. However, $N_0$ cannot set the \emph{second-next-hop} field in $P_{0}$ because $N_0$ does not know the \emph{second-next-hop} from the intended forwarder's point of view (i.e., the\emph{ second-next-hop} from $F_0$). Now, if $F_{1}$ receives and decodes $P_{0}$ successfully, it can consult the routing module and find the next-hop because $F_{1}$ is the designated intended forwarder. However, if non-intended forwarder $N_{1}$ receives the coded packet, since \emph{second-next-hop} was not set and also $N_1$ was not specified in the route, it may not be able to find the correct next-hop. Thus, $N_{1}$ as a non-intended forwarder must drop coded packets.

A preliminary version of FlexONC~\cite{FlexONC-Kafaie-ICC2015} moves one step further for more diffusion gain than BEND, and allows non-intended forwarders to cooperate in receiving and forwarding not only native packets but also coded packets. In fact, it provides the next-hop information of decoded packets to non-intended forwarders so that they are able to forward the packet to the correct next-hop toward the destination. As we explained in the previous section, by doing so, FlexONC provides more \emph{diffusion gain} and more coding opportunities, which lead to a higher throughput in comparison to previous methods. 

In this article, we discover and address the problem related to the \emph{common coding conditions}, and we augment the implementation of FlexONC to incorporate our solution for this problem. In addition, we further discuss FlexONC as a Media Access Control (MAC) layer solution that not only increases the coding opportunities in the network, but also allows to control effectively how far packets stray away from a designated shortest path. We conduct more experiments to show the efficiency of our solution by comparing FlexONC with other schemes from different aspects such as throughput, end-to-end delay, the number of duplicate packets, the number of coding opportunities, and overall overhead and complexity.

\section{FlexONC – Motivating Examples} \label{section:examples}

\subsection{More Diffusion Gain}
\begin{figure}[ht]
\centering
\includegraphics[scale=0.33]{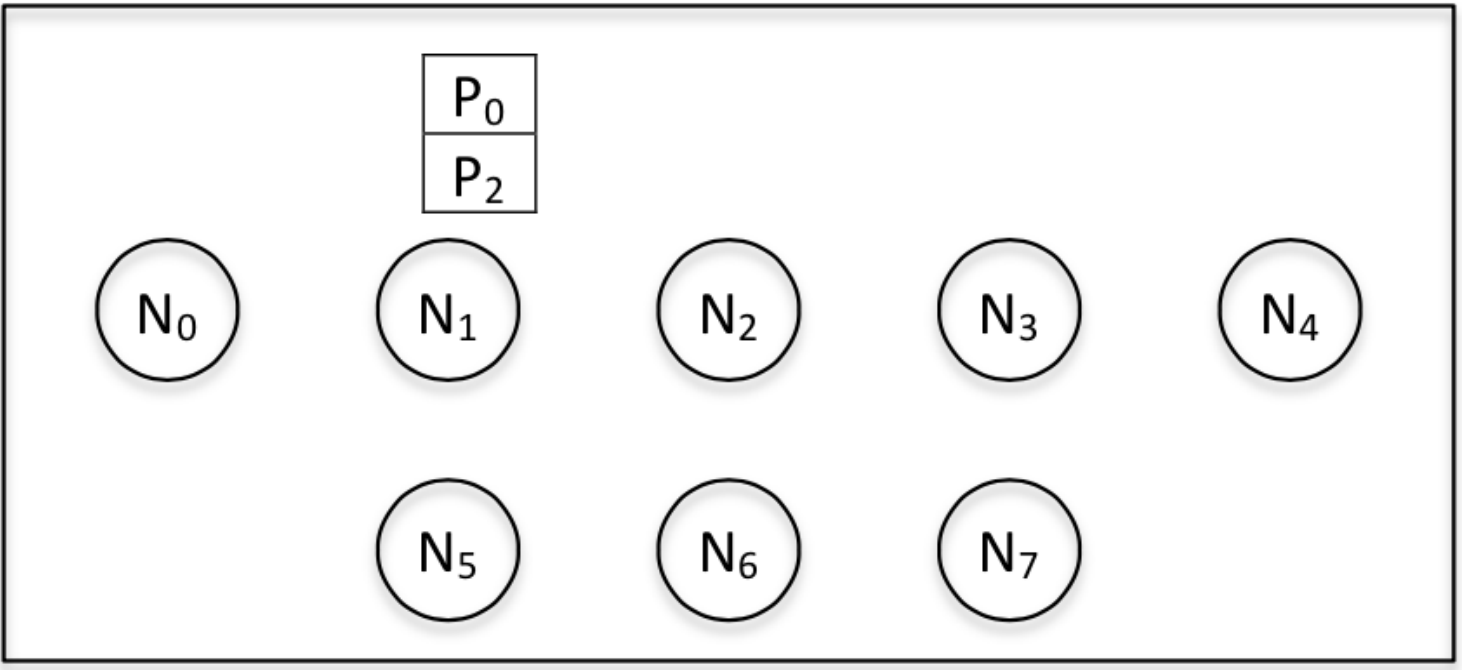}
\caption{Non-intended forwarders can help decoding.}
\label{figure:8node}
\end{figure}

\figurename~\ref{figure:8node} presents an 8-node topology where there exist two flows from $N_{0}$ to $N_{4}$, and vice versa. In all topologies used in this research, we assume each node can receive packets only from nodes immediately next to it horizontally, vertically, or diagonally. As shown in this figure, $N_{1}$'s queue contains 2 native packets $P_{0}$ and $P_{2}$ with different next-hops $N_{0}$ and $N_{2}$, respectively. Let us assume $P_{0}$'s next-hop is $P_{2}$'s previous forwarder or one of its neighbors, and vice versa. So, $N_{1}$ decides to mix these packets together, hoping that $N_{2}$ ($N_{0}$) has already received $P_{0}$ ($P_{2}$) and it can decode $P_{2}$ ($P_{0}$). Therefore, $N_{1}$ sends a coded packet $P=P_{0} \oplus P_{2}$ to $N_{0}$ and $N_{2}$ (i.e., next-hop list in the packet header contains $N_{0}$ and $N_{2}$) while we assume $N_{6}$ overhears the packet. 

In the previous methods like COPE and BEND, $N_{6}$ discards the packet immediately because either it is not the next-hop (as in COPE) or the packet is not a native packet (as in BEND). Here, we assume that $N_{2}$ does not receive the coded packet or $P_{0}$, so it cannot decode $P_{2}$, and that $N_{6}$ receives it successfully, and also can decode the packet. In such a scenario, in previous methods, after a time-out, $N_{1}$, which has not heard any ACK from $N_{2}$, retransmits the packet. However, FlexONC avoids such unnecessary retransmissions, and $N_{6}$ forwards the packet to its next-hop on behalf of $N_{2}$.

In fact, FlexONC allows non-intended forwarders like $N_{6}$ to decode a received coded packet if they can, and forward it toward the final destination as long as the intended forwarder fails to do so. By doing so, since $N_{2}$ is not the only node in charge of forwarding packets, the traffic is spread in the network. That is if $N_{2}$ fails to receive or decode a packet, its role is immediately covered by $N_{6}$. This idea not only can accelerate packet delivery by removing some retransmissions but also can provide more coding opportunities. For example, let us further assume $N_{6}$ is going to forward $P_{2}$ on behalf of $N_{2}$. If $P_{2}$ is eligible to be mixed with some packets queued at $N_{6}$, by allowing $N_{6}$ to decode and forward it, we capture more coding opportunities in $N_{6}$. However as will be described later, we provide some strategies to ensure that the nodes do not stray far away from the original route, and also to limit the number of duplicate packets in the network.

\subsection{ Right Coding Opportunities} \label{EXcodingIssue}
\begin{figure}[ht]
\centering
\includegraphics[scale=0.33]{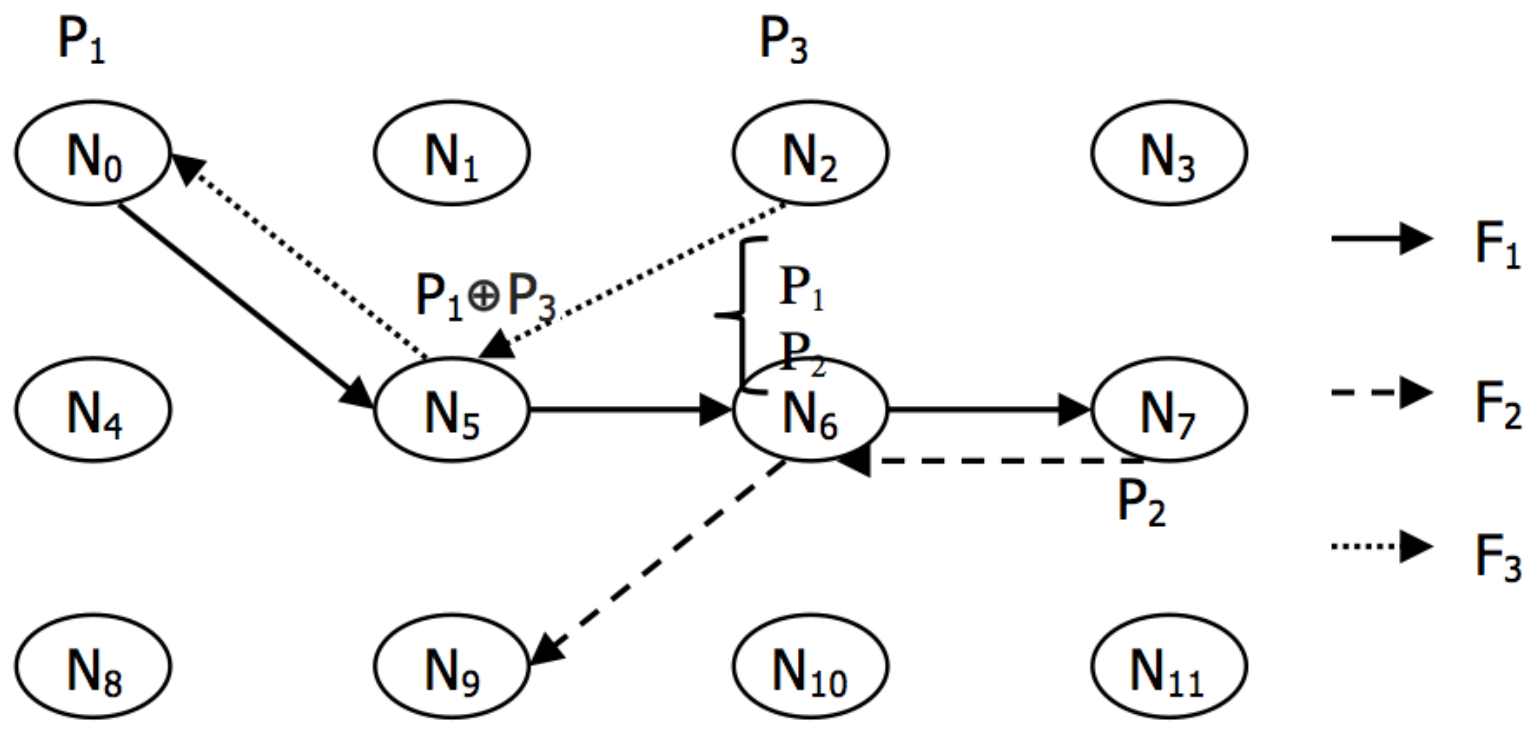}
\caption{Common coding conditions are not sufficient.}
\label{figure:12node}
\end{figure}

Let us assume that in the grid topology provided in \figurename~\ref{figure:12node}, our focus is on three specific flows: 1) $F_1$ with packets like $P_1$ from $N_0$ to $N_7$, 2) $F_2$ with packets like $P_2$ from $N_7$ to $N_9$, and 3) $F_3$ with packets like $P_3$ from $N_2$ to $N_0$. Let us further assume that $N_5$ transmits a coded packet from flows $F_1$ and $F_3$, $P_1 \oplus P_3$. We assume $N_6$, as the intended forwarder of $P_1$ can decode the packet successfully, but $N_9$ cannot decode it as $N_9$ cannot overhear $P_3$.  Let us call a packet like $P_1$, which has been received coded by the node and then it is decoded, a \emph{decoded-native} packet. Now, the question is that under what conditions a node (e.g., $N_6$) can combine a decoded-native packet (e.g., $P_1$) with other packets? For example, can $N_6$ combine packets received from $N_5$ and $N_7$? Are the common coding conditions enough to decide on encoding such packets?


Based on the common coding conditions, the combination of $P_1$ and $P_2$ at $N_6$ seems a valid encoding strategy because the next-hop of $P_1$ (i.e., $N_7$) is the previous hop of $P_2$, and the next-hop of $P_2$ (i.e., $N_9$) is one of the neighbors of the previous hop of $P_1$ (i.e., $N_5$). However, one may notice that if $N_9$ receives the coded packet $P_1 \oplus P_2$, it cannot decode $P_2$ correctly as it has only overheard  $P_1 \oplus P_3$ and neither $P_1$ nor $P_3$. In fact, the problem happens because the previous hop of $P_1$ (i.e., $N_5$) sends it as a coded packet; therefore its neighbors (e.g., $N_9$) do not receive $P_1$ natively. As a result, if $N_6$ encodes this decoded-native packet, $N_9$ cannot decode the received coded packet $P_1 \oplus P_2$. 

Note that although COPE uses reception reports, in such a scenario COPE could not rely on them for encoding. Since $N_9$ does not send any packet, it has to send the reception reports periodically, which reduces the probability that its neighbors receive a fresh report on time. Therefore, most of the time the neighbors do not have deterministic information required for encoding and would need to guess based on the delivery probability between nodes. Hence, if the delivery probability between different nodes is high, in COPE, $N_6$ will encode $P_1$ and $P_2$. To show the severity of the issue, we ran simulations, using a simulation version of COPE in ns-2, to decide on encoding of the packets in the topology depicted in \figurename~\ref{figure:12node}.

\begin{figure}[ht]
\centering
\includegraphics[scale=0.4]{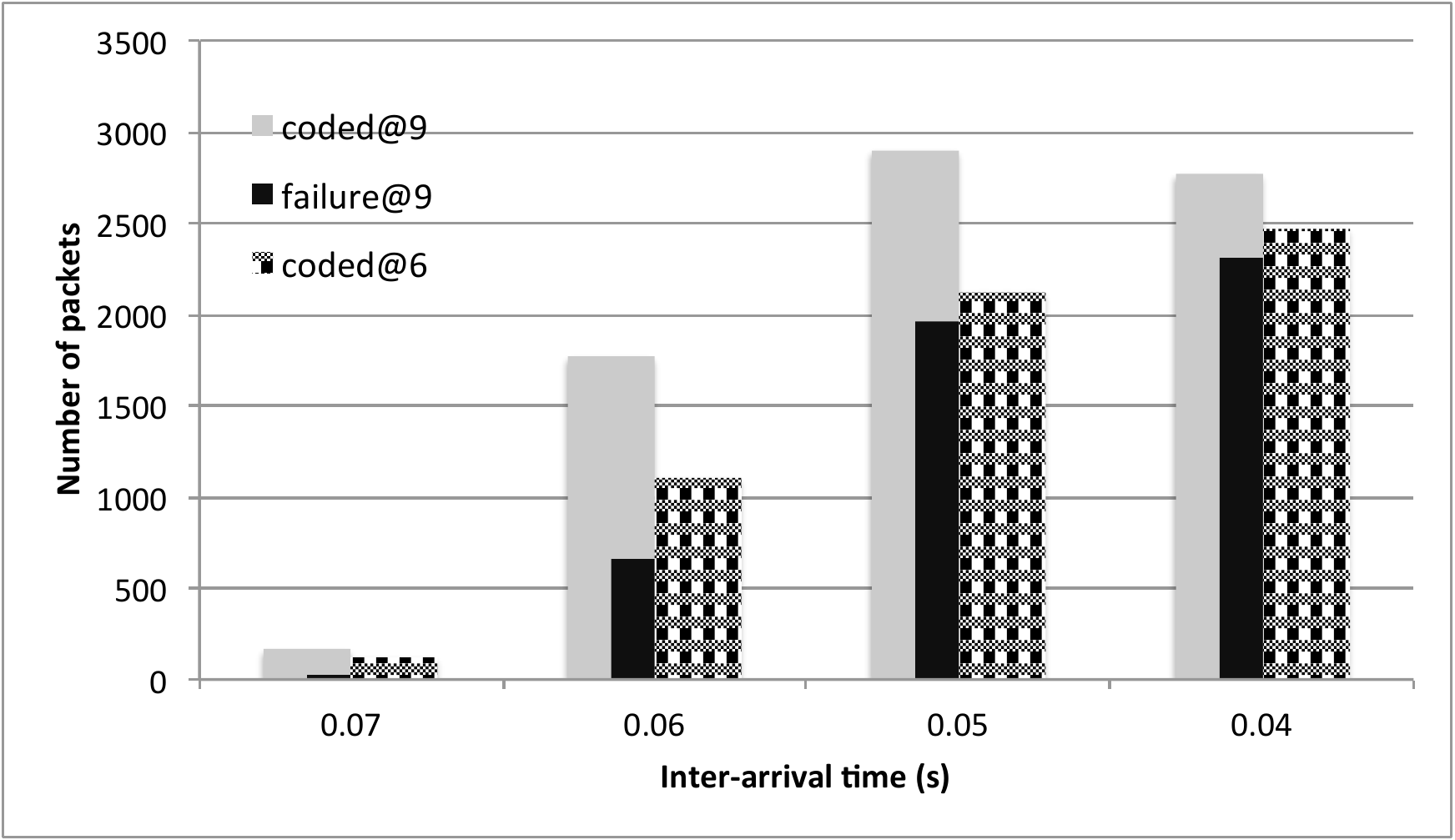}
\caption{Decoding failure of COPE by applying the common coding conditions.}
\label{figure:COPEissue}
\end{figure}

\figurename~\ref{figure:COPEissue} presents the number of coded packets received by $N_6$ (i.e., coded@6), the number of coded packets received by $N_9$ (i.e., coded@9), and also the number of coded packets that $N_9$ cannot decode (i.e., failure@9) because of the explained issue. As shown in this figure, by decreasing the inter-arrival time (i.e., increasing the arrival rate), the length of the transmission queue as well as the coding opportunities at nodes increase. Therefore, the probability that an encoded packet received and decoded by $N_6$ (i.e., a decoded-native packet) can be encoded again increases, which in this scenario causes the explained issue and consequently increases decoding failures at $N_9$. 

This example and simulation results show that the common coding conditions are not enough, and more restrictive coding conditions are required to address the issue stated here. Therefore, we not only provide cooperative forwarding for native and coded packets, but also address this issue by proposing an additional rule to restrict the common coding conditions.

\section{Design Details} \label{section:design}
As described earlier, the idea behind FlexONC is to have backup nodes to decode and forward a packet in case where the intended forwarder fails, either due to unsuccessful reception of the packet or lack of required packets in the buffer to decode the original packet. In addition, FlexONC provides more comprehensive coding conditions and a mechanism to detect right coding opportunities and avoid undecodable encodings. In this section, we first discuss some of the challenges that FlexONC addresses. Then, we describe in detail the responsibilities of the sender and receiver of a coded packet to realize these ideas, and address these challenges.

\subsection{Objective and Challenges}
FlexONC should avoid unnecessary changes to the standard MAC protocols, and be as simple as possible to be feasible in real scenarios. Moreover, it should be compatible with different routing protocols despite few modifications. To realize such compatibility, while having more flexibility and accuracy in forwarding and coding, FlexONC should address the following questions. 
\begin{itemize}
\item How to select the nodes that can help the intended forwarder to forward packets: In other words, how should we decide which nodes are eligible for packet forwarding? For example, in \figurename~\ref{figure:8node}, when $N_{1}$ sends the packet, $N_{5}$, $N_{2}$ and $N_{6}$ may receive it, but are they good candidates to forward the packet? 
\item How to limit the number of duplicate packets: Since more nodes cooperate to move packets toward the destination, their imperfect collaboration may cause a significant number of duplicate packets travelling in the network leading to unnecessary contention and collision. Some mechanisms are required to control duplicate packets. 
\item How to provide flexible forwarding but not too far from the specified route: Although in FlexONC, like BEND, packets may not follow the exact route specified by the routing protocol, we need to keep them around the determined route. To do so, BEND uses the \emph{second-next-hop} field in native packets. However, as we described earlier, it is not applicable to coded packets at non-intended forwarders. For example, in \figurename~\ref{figure:8node} when $N_{6}$ receives the coded packet, even if it can decode $P_{2}$, it does not know the address of the next-hop from $N_{2}$ toward the destination. Thus in FlexONC, we need a new approach for non-intended forwarders to find the correct address of the next-hop.
\item How to propose a complete set of rules to combine packets: As illustrated in Subsection~\ref{EXcodingIssue}, the common coding conditions used in other inter-flow network coding methods are not accurate enough to recognize right coding opportunities in some scenarios, and may lead to decoding failures. The question is how to establish a complete set of rules to correctly decide on mixing the packets of flows which are decodable at the next-hop?
\end{itemize}
We address all these aspects in the next subsections.

\subsection{Decoding and Forwarding Strategy}
In FlexONC, nodes in the network are in promiscuous mode, and store all received and overheard packets in a buffer, called \emph{coding buffer}. Each packet is kept there for a period of time, long enough that the node can use these packets to decode the received coded packets. In case of successful decoding, the receiver sends an ACK while a NACK (i.e., negative acknowledgement) signals failure in decoding. 
In terms of forwarding, native packets are only sent by intended forwarders. A non-intended forwarder may forward a packet on behalf of an intended forwarder if the non-intended forwarder can provide more coding opportunities.

In FlexONC, although packets may not follow the exact route specified by the routing protocol, they travel near it and do not stray too far away. Thus, when a non-intended forwarder forwards the packet on behalf of the intended forwarder, it should send it to the next-hop toward the destination from the intended forwarder's point of view. For example in \figurename~\ref{figure:8node}, when $N_{1}$ sends the coded packet $P=P_{0} \oplus P_{2}$, $N_{0}$, $N_{5}$, $N_{2}$, and $N_{6}$ may receive the packet. If $N_{2}$, which is the intended next-hop for $P_{2}$, fails to receive the packet successfully, and if one of the non-intended forwarders (e.g., $N_{5}$, $N_{0}$, $N_{6}$) wants to forward it, they need to know the address of the next-hop from $N_{2}$ toward the destination (not from themselves), which is $N_{3}$ in this example.

Since the \emph{second-next-hop} field in BEND cannot solve this problem, instead of adding this field to the packet header, in FlexONC, the routing protocol is enhanced such that each node also maintains forwarding tables of all its neighbors. As such, when for example $N_{6}$ forwards $P_{2}$ on behalf of $N_{2}$, it knows the address of the next-hop from $N_{2}$ toward the destination, and simply sends the packet to it.

\subsection{Receivers in FlexONC}
Since every node in the vicinity of the sender can receive the packet, we classify the receivers of a packet in two groups, intended forwarders and non-intended forwarders. As summarized in Table~\ref{table:definition}, an intended forwarder is a node whose address has been specified in the packet header as the next-hop of the packet by the routing protocol. On the other hand, non-intended forwarders are the nodes that are in the neighborhood of the next-hop and can help it in forwarding packets.

When a sender transmits a coded packet, all of its neighbors may receive it. However, every node that receives the packet is not necessarily eligible to forward it. In addition, if all eligible nodes were to forward the same packet, that would be a waste of the network bandwidth as well as a source of collision. We need a method to choose and prioritize eligible forwarders.

A node is an \emph{eligible} non-intended forwarder if it is not only the neighbor of the sender but also a neighbor of both next-hop and the second next-hop of a coding partner. Following this rule ensures that a packet would travel correctly toward its final destination, even if it is forwarded by a different node than its next-hop. In the rest of the paper, we use the term ``non-intended forwarder'' to refer to ``eligible non-intended forwarders''. 

If an intended forwarder (e.g., $N_{2}$ in \figurename~\ref{figure:8node}) receives a coded packet and can decode the packet, it simply replies with an ACK. However, if it cannot decode the packet, it sends a NACK instead. In FlexONC, ACKs and NACKs contain the address of their sender (i.e., the transmitter of ACK/NACK) instead of the receiver, the same as in BEND. If non-intended forwarders (e.g., $N_{6}$) hear the ACK, they realize that the intended forwarder has decoded the packet successfully and does not need their help.

In FlexONC, when a node like $N_{6}$ in \figurename~\ref{figure:8node} receives a coded packet, it first looks for its address in the next-hop list. If it cannot find its address, clearly it is not the intended forwarder for any coding partner in the coded packet. Therefore, $N_{6}$ searches for a native packet in the coded packet that 1) its intended forwarder (e.g., $N_{2}$ for $P_{2}$ in \figurename~\ref{figure:8node}) is $N_{6}$'s neighbor, 2) its next-hop from the intended forwarder (e.g., $N_{3}$ for $P_{2}$ in \figurename~\ref{figure:8node}) is $N_{6}$'s neighbor, and 3) it is decodable by $N_{6}$. Based on these criteria, in \figurename~\ref{figure:8node}, although when $N_{1}$ sends the coded packet $P$, $N_{0}$, $N_{5}$ and $N_{6}$ as well as $N_{2}$ may receive the packet, $N_{0}$ is not eligible to forward $P_{2}$ due to the first criterion. Furthermore, $N_{5}$ is not qualified for the second criterion, and therefore $N_{6}$ is the only non-intended forwarder which can send $P_{2}$ on behalf of $N_{2}$ if it can decode it.

\begin{figure}[ht]
\centering
\includegraphics[scale=0.5]{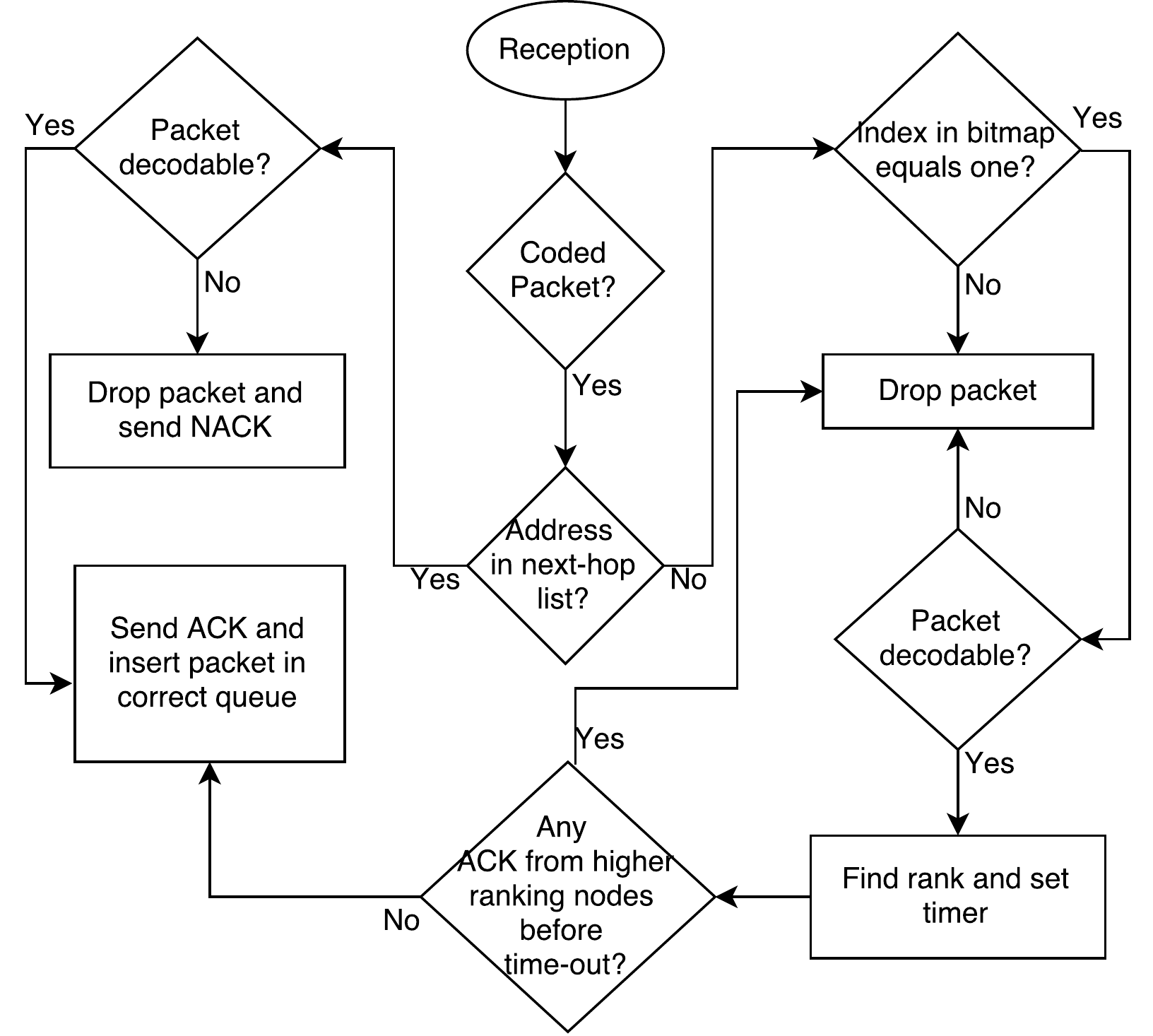}
\caption{Flowchart for receivers of coded packets in FlexONC.}
\label{figure:flowChart}
\end{figure}

However, a non-intended forwarder should not forward a packet immediately after decoding it because the intended forwarder may forward the packet itself and would not need the non-intended forwarders' help. In addition, if there are more than one eligible non-intended forwarder, an ordering among them is required to avoid the transmission of more than one ACK to the packet sender. Due to this reason, in FlexONC the sender adds the index of all eligible non-intended forwarders to the packet header.\footnote{We assume that all nodes in the network agree on the same numbering system which represents each of them with a unique index known by all other nodes.} Specifically, when a non-intended forwarder receives a coded packet, it sorts the list of indexes (i.e., all non-intended forwarders), gives the first priority to the intended forwarder of the decoded packet, and considers its index in the sorted list as its rank. Then, it sets a timer and waits for an ACK from any node with a higher rank. If it does not hear any ACK after time-out, it is likely that none of the nodes with a higher rank has received and can forward the packet, so it is its turn to send the ACK back to the sender, mixes possibly the decoded packet with other packets in the queue, and forwards it. \figurename~\ref{figure:flowChart} presents the flowchart for receivers of a coded packet in FlexONC.

\subsection{Senders in FlexONC}
When a node sends a coded packet, it adds the list of the next-hops of all coding partners to the packet header. Thus when each next-hop receives the packet, it does not send the acknowledgement (either ACK or NACK) immediately but after some time proportional to its position in the next-hop list as well as the transmission and propagation time of the acknowledgement. For example, if a node transmits the combination of 3 packets with the next-hops $N_1$, $N_2$ and $N_3$, after receiving the coded packet, $N_3$ waits for a certain amount of time to ensure that $N_1$ and $N_2$ have sent their packet acknowledgements, and then $N_3$ sends back ACK/NACK.

Furthermore, the sender detects all eligible non-intended forwarders of a coded packet, and adds a bitmap to the packet header where each bit represents one of the nodes in the network (as discussed in Subsection~\ref{subsection:overall}, the overhead introduced by adding this bitmap is less than a few bytes). If the node is an eligible forwarder, the corresponding bit is set to $1$, otherwise the bit keeps the default value which is $0$. We assume that each node is represented with a unique index known by all other nodes, and each node ranks eligible non-intended forwarders based on their indexes.

\begin{figure}[ht]
\centering
\includegraphics[scale=0.56]{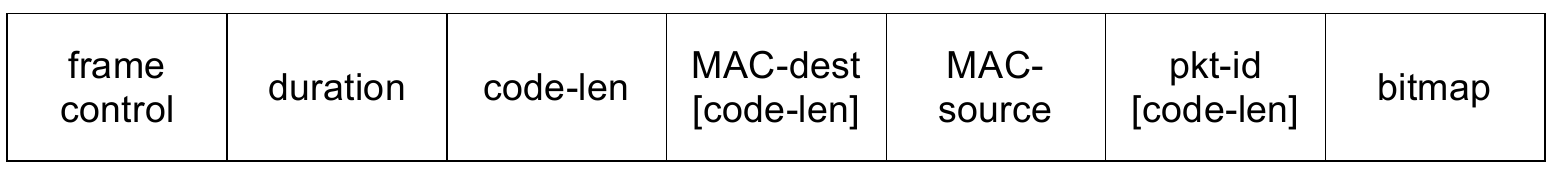}
\caption{MAC header for coded packets.}
\label{figure:codedMACheader}
\end{figure}

In FlexONC, the fields in the packet header of native packets do not change. However, the MAC-layer header of coded packets includes some additional information, such as the number of coding partners, the bitmap, and the address of the next hop and the packet-id of all coding partners as presented in \figurename~\ref{figure:codedMACheader}. Note that we keep the original format of the upper layers' headers, and the \emph{XOR} of the coding partners is added to the MAC data-frame as payload.

Since the sender stores the forwarding table of its neighbors, it can check which neighbors are eligible non-intended forwarders. Doing so, the sender can calculate its maximum waiting time for receiving an ACK which is proportional to the number of the next-hops (i.e., intended forwarders) and eligible non-intended forwarders of coding partners. It is obvious that when a sender sends a combination of $n$ packets, it should wait to receive $n$ ACKs. Thus, its waiting time before time-out is more than when it transmits a native packet. In FlexONC, because more nodes can help in decoding and forwarding a packet, if the sender does not hear an ACK from the intended forwarder, there is still a chance that it receives the ACK from a non-intended forwarder. Therefore, the sender should wait a little longer before it retransmits the packet. As such, in FlexONC the waiting time of the sender for coded packets is calculated in terms of the number of both coding partners and eligible non-intended forwarders.

\begin{figure}[ht]
\centering
\includegraphics[scale=0.65]{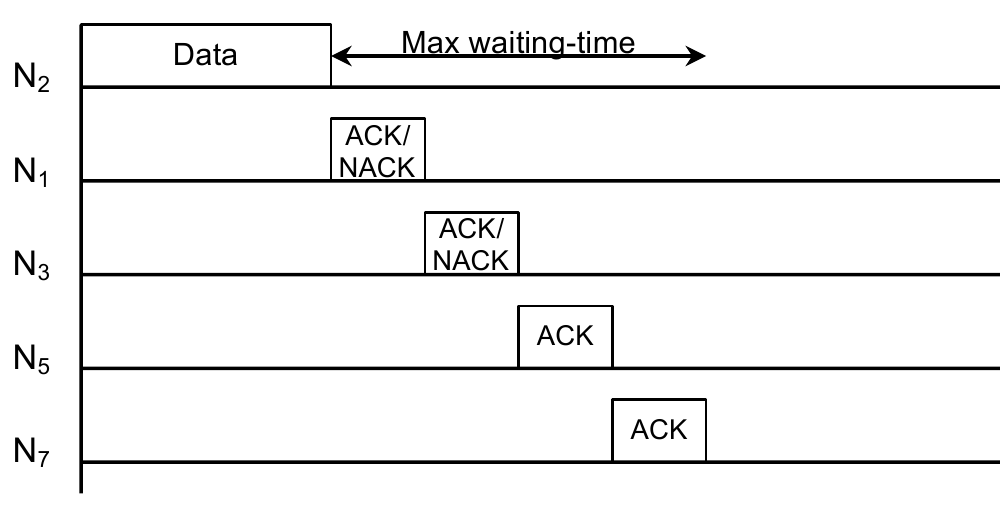}
\caption{The time-window dedicated to different nodes to send back the acknowledgment, where in the topology depicted in \figurename~\ref{figure:8node} $N_2$ transmits a coded packet to the next-hops $N_1$ and $N_3$, and $N_5$ and $N_7$ are non-intended forwarders.}
\label{figure:timeWindow}
\end{figure}

To illustrate the idea in more details, let us assume that in \figurename~\ref{figure:8node}, $N_2$ mixes two native packets and forwards the coded packet to the next-hops $N_1$ and $N_3$ (i.e., $N_1$ and $N_3$ are the intended forwarders of these two packets), while $N_5$ and $N_7$ are eligible non-intended forwarders specified in the bitmap. \figurename~\ref{figure:timeWindow} shows the maximum waiting time at the sender, $N_2$, after transmitting the data packet and the time-window dedicated to the intended and non-intended forwarders to reply if they need. Note that the intended forwarders reply by an ACK after successful decoding and send a NACK after decoding failure. In addition, a non-intended forwarder replies by an ACK only if decoding is successful and no ACK was heard from neither the corresponding intended forwarder nor higher-ranking non-intended forwarders. 

When the sender receives an ACK for a packet, it removes the packet from its transmission queue; it may still keep it in the coding buffer for decoding purposes. On the other hand, when the sender receives a NACK for the sent packet, it keeps waiting until either time-out or receiving an ACK for the same packet. In the case of time-out for native packets, the sender retransmits the same packet if the number of transmissions does not exceed the maximum retransmission count. However, for coded packets, if the node receives ACKs or NACKs for none of the coding partners, it retransmits the same coded packet. Otherwise, it inserts the coding partners which are not ACKed in the transmission queue. 

\subsection{Encoding Strategy}
As explained earlier to decide on encoding packets, the majority of encoding methods, within a two-hop region, use a similar coding structure called \emph{two-hop coding structure}~\cite{XNCSurvey-Xie-ComNet2015} with the same coding conditions~\cite{COPE-Katti-IEEEACMTransactions2008, BEND-Zhang-CNJournal2010, InterFlow-Huang-PIMRC11, NCAQM-Seferoglu-NetCod10, BRONC-Guo-CCPR10, NCDS-Wang-INFOCOM2013}. Based on these \emph{common coding conditions}, node $N$ can combine two packets $P_1$ and $P_2$ if:
\begin{enumerate}
\item The next-hop of $P_1$ is the previous hop of $P_2$ or one of its neighbors.
\item The next-hop of $P_2$ is the previous hop of $P_1$ or one of its neighbors.
\end{enumerate}

However, as illustrated in Subsection~\ref{EXcodingIssue} in some scenarios such as \figurename~\ref{figure:12node}, these coding conditions are not sufficient. In fact, the issue happens because in the common coding conditions, it is assumed that all the neighbors of the previous hop (e.g., $N_5$) are able to decode the coded packet sent by it (e.g., $P_1 \oplus P_3$). However, this is not necessarily a valid assumption as some of these neighbors (e.g., $N_9$) may not be able to do so. To address this issue, we add an additional condition to the common coding conditions as follows.

\emph{RecodingRule} - To combine a decoded-native packet (i.e.,  a packet received as a coded packet from its previous hop and has been decoded) with other packets (i.e., recode the packet), the node does not check the neighborhood of the previous hop of the packet. In fact, if $P_1$ is a decoded-native packet the common coding conditions should be modified as follows:
\begin{enumerate}
\item The next-hop of $P_1$ is the previous hop of $P_2$ or one of its neighbors.
\item The next-hop of $P_2$ is the previous hop of $P_1$.
\end{enumerate}

\begin{figure}[ht]
\centering
\includegraphics[scale=0.4]{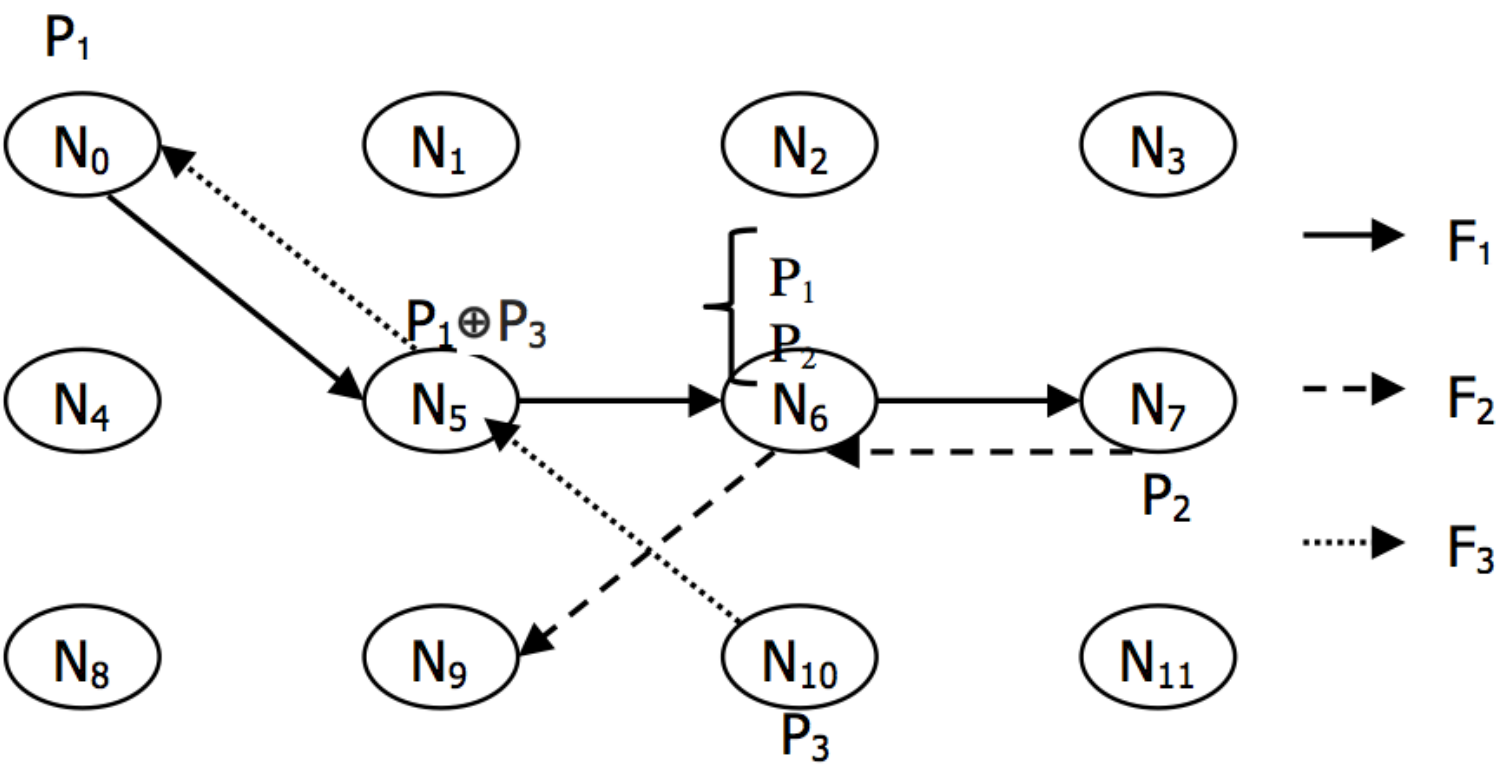}
\caption{RecodingRule, sufficient but not necessary.}
\label{figure:12node2}
\end{figure}

\emph{RecodingRule} is sufficient but may not always be necessary. That is, although it avoids misleading coding opportunities and decoding failures in the scenario depicted in \figurename~\ref{figure:12node}, in some other scenarios it limits the number of right coding opportunities in the network. As an example, let us describe the effect of our \emph{RecodingRule} on the scenario presented in \figurename~\ref{figure:12node2}. In this figure, the route of flow $F_3$, in comparison to \figurename~\ref{figure:12node}, has changed  so that $N_9$ can overhear the packets of this flow. Now, $N_9$ overhears $P_3$ from $N_{10}$, and $P_1 \oplus P_3$ from $N_5$. As a result, we do not need to apply \emph{RecodingRule}, and $N_9$ can decode $P_1 \oplus P_2$ received from $N_6$ successfully.

Therefore, \emph{RecodingRule} should be intelligently used only in cases that the interaction between flows is so that the common coding conditions may provide misleading coding opportunities. This type of encoded packets cannot be decoded in the next-hop, and the sender will receive a NACK for it. Thus, we propose a solution called \emph{SwitchRule} to decide properly on applying \emph{RecodingRule} on different flows at different nodes. In fact, \emph{SwitchRule}, based on the received NACKs for each flow at each node, decides to switch back and forth to use and not to use \emph{RecodingRule}. Note that  \emph{SwitchRule} only needs to be applied at the flow-granularity, not the packet-granularity. 

At the beginning, every node uses the common coding conditions to encode packets. However, when each node combines a decoded-native packet, $P_1$, with another packet, $P_2$, if the next-hop of $P_2$ is not the previous hop of $P_1$ but one of its neighbors, $P_1$ is tagged as a \emph{suspect} packet. This means we are suspicious that decoding failure may happen because the next-hop of $P_1$'s partner (i.e., $P_2$) may have not overheard the suspect packet, $P_1$. Each node keeps track of the number of NACKs received for the partners of suspect packets of each flow. If the number of NACKS for a flow is greater than a threshold, the node applies \emph{RecodingRule} for the rest of the packets of that particular flow. This means the node will not combine a decoded-native packet of that flow with any other partner if the next-hop of the partner is not the previous-hop of the decoded-native packet.

\begin{figure}[ht]
\centering
\includegraphics[scale=0.6]{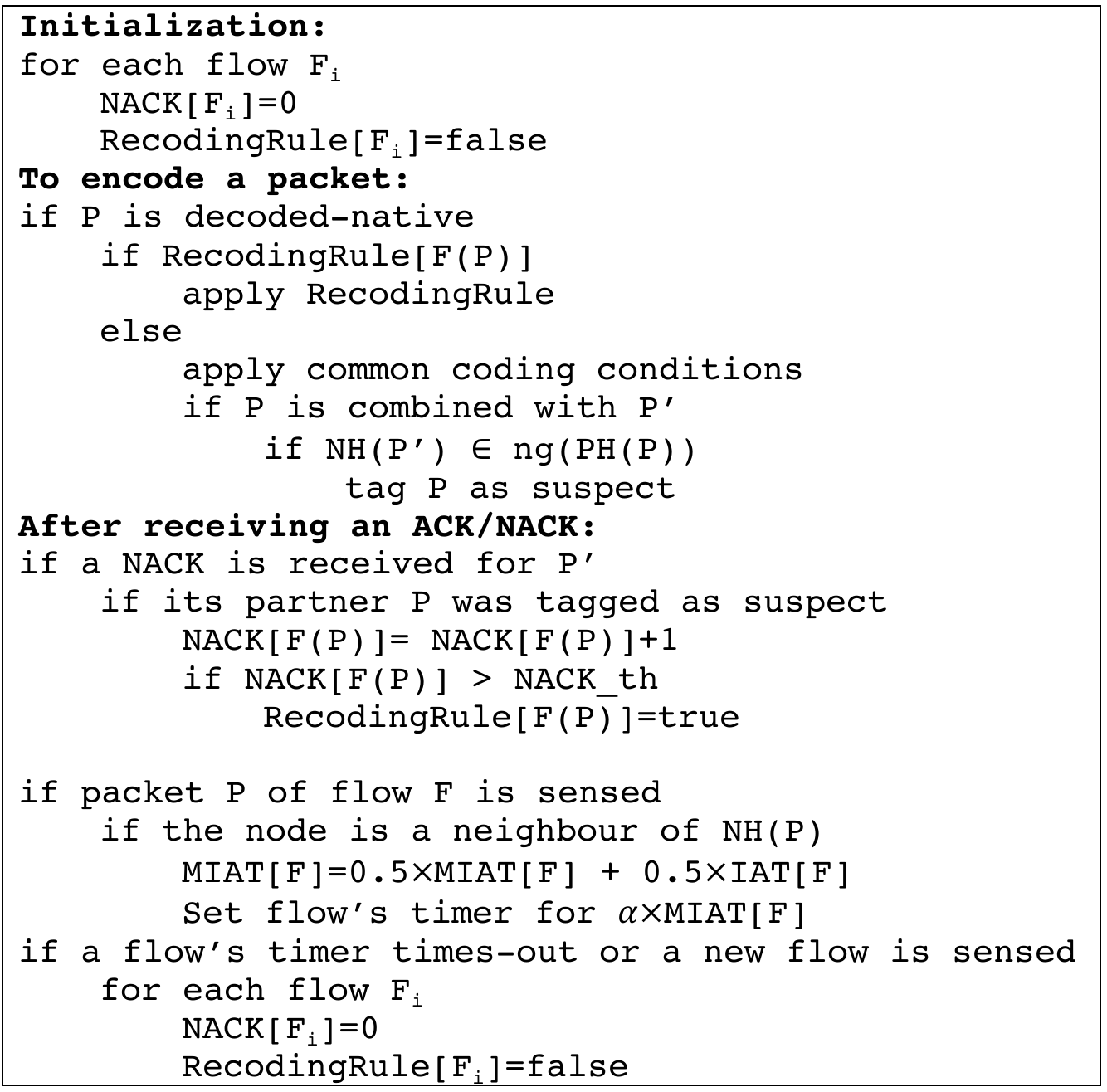}
\caption{Pseudo-code of \emph{SwitchRule}. The number of NACKs received for flow F is stored in NACK[F]. NH(P), PH(P) and F(P) denote the next-hop, the previous-hop and the flow of P, respectively. ng(N) represents the set of neighbors of node N. IAT[F] and MIAT[F] denote the inter-arrival time and the mean inter-arrival time of flow F. The timer for flow F is set to $\alpha$ times of MIAT[F], where $\alpha>1$.}
\label{figure:encodingPseudo}
\end{figure}

Furthermore, a node will switch back to not using \emph{RecodingRule} whenever it hears packets of a new flow or it does not hear any packet from a flow anymore. To implement the latter case in \emph{SwitchRule}, each node set a timer for each flow. If the timer of a flow times-out before receiving a new packet of that flow, the node switches back to the common coding conditions for all flows. The waiting time before the time-out is several times of the estimated inter-arrival time of the packets of the flow. The inter-arrival time of each flow is estimated using a weighted-average over the previous average and the latest measured inter-arrival time. \figurename~\ref{figure:encodingPseudo} presents the pseudo-code of the \emph{SwitchRule}'s mechanism.

\subsection{How to Limit the Number of Duplicate Packets?} \label{duplicate}
Although FlexONC aims to eliminate duplicate packets by prioritizing non-intended forwarders and making the sender wait for their ACKs, duplicate packets may still exist in the network, due to various reasons such as lack of perfect synchronization. For example, a non-intended forwarder may not hear the ACK sent by the intended forwarder or higher-ranking non-intended forwarders, and transmit the packet unnecessarily. Therefore, FlexONC relies on more strategies to control the number of duplicate packets in the network. 

First, after receiving an ACK for a given packet-id, if the node finds a packet with the same packet-id in its transmission queue that the sender of the ACK is the next-hop of the packet or one of corresponding eligible non-intended forwarders, the node drops the packet (i.e., the packet has already been received by down stream nodes). Second, in FlexONC each node stores a limited number of received ACKs, and if it receives a packet, it searches this ACK list. If it finds an ACK for the same packet sent by its next-hop or one of its eligible non-intended forwarders, it also drops the packet.

\section{Performance Evaluation} \label{section:performance}

\begin{table}[!t]
\renewcommand{\arraystretch}{1.3}
\caption{Information available at nodes in different schemes.}
\label{table:nodeInformation}
\centering
\begin{tabular}{|c||c||c||c||c||c|}
\hline
 \textbf{Information} & \textbf{Non-} & \textbf{COPE} & \textbf{CORE} & \textbf{BEND} & \textbf{FlexONC}\\
 & \textbf{coding} & & & & \\
 \hline
next-hop & $\surd$ & $\surd$& & $\surd$ &$\surd$ \\  \hline
second & & & & \multirow{2}{*}{$\surd$} &  \\  
next-hop & & & & & \\ \hline
neighbors'  & & & & & \multirow{2}{*}{$\surd$} \\ 
forwarding info & & & & &\\  \hline
forwarder set & & &$\surd$  & & $\surd$  \\  \hline
node's & & & \multirow{2}{*}{$\surd$} & & \\ \
geo-position & & & & & \\ \hline
\end{tabular}
\end{table}

We use the Network Simulator (ns-2) to compare the performance of FlexONC, with and without \emph{RecodingRule}, against the non-coding scheme, a simulation version of COPE as a prominent research on network coding, and two opportunisitc forwarding schemes in network coding (i.e., BEND and CORE).\footnote{Note that in all simulations, IEEE 802.11~\cite{IEEE802.11-2007} is selected as the data link layer signaling method.} Table~\ref{table:nodeInformation} summarizes the type of information provided at nodes in different schemes. The rest of this section describes the experiment scenarios as well as the performance results in three different topologies.

\subsection{Settings}
To study the performance under different link qualities and packet loss probabilities in our simulation, bit error rate (BER) is added to the physical layer. In fact, even if the signal strength of a received packet is higher than reception threshold, the packet may still be dropped with a probability calculated in terms of BER. BEND and CORE also use a similar physical layer model. The channel propagation used in ns-2 is a two-ray ground reflection model~\cite{TwoRay-Rappaport}, and the maximum transmission range is $250$ m. The data rate is fixed to 1 Mbps. The sources, in our simulation scenarios, send CBR (constant bit rate) data flows with a datagram size of $1000$ bytes. Also, we use DSDV (Destination-Sequenced Distance-Vector)~\cite{DSDV-Perkins-1994} as the routing protocol and apply a few minor changes so that each node can obtain forwarding tables from its neighbors.

We compare the performance of FlexONC with other baselines in several scenarios. In the first part, we use scenarios in which common coding conditions are enough to encode the packets in all methods, including FlexONC. Then in the second part, we present the performance of different methods in scenarios where \emph{RecodingRule} is required to avoid erroneous encoding causing decoding failures. 
 
\subsection{Performance under Common Coding Conditions}
To investigate the performance of FlexONC in comparison to BEND, CORE, COPE and the non-coding scheme, we test them in different scenarios and compare their throughput as well as the throughput gain of FlexONC over the baselines for different BERs in two topologies. First, we compare them using a simple 8-node topology shown in \figurename~\ref{figure:8node}, and then we use a $5 \times 5$ grid topology as a more general case. In both topologies, different flows have been selected so that in most cases the common coding conditions are enough and we compare all methods using the same coding conditions (i.e., common coding conditions).

\subsubsection{8-Node Topology} \label{subsection:8node}
In the 8-node topology presented in \figurename~\ref{figure:8node}, two flows in opposite directions transmit packets from $N_{0}$ to $N_{4}$ and vice versa. Since the distance between adjacent nodes in both $X$ and $Y$ axes is 150 m, each node can receive packets only from nodes immediately next to it horizontally, vertically, or diagonally (e.g., $N_{1}$ can hear from $N_{0}$, $N_{5}$, $N_{2}$, and $N_{6}$). The inter-arrival time of CBR flows in these scenarios is 0.07 s and its duration is 150 s. 

\begin{figure}[!t]
\centering
\includegraphics[scale=0.6]{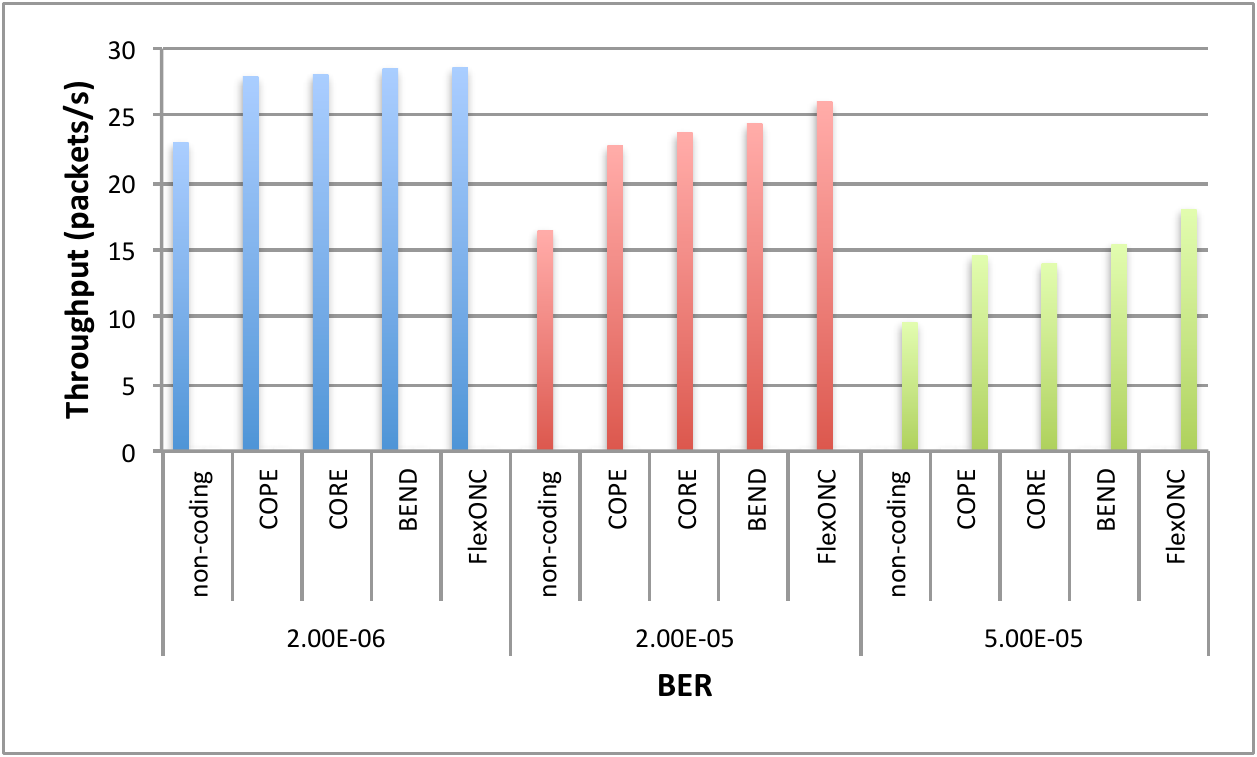} 
\caption {Throughput of different methods in 8-node topology for different BERs.}
\label{figure:throughput8}
\end{figure}

In this topology, for each intended forwarder except for the destination, there exists at least one non-intended forwarder that can help the intended forwarder and forward packets when the intended forwarder fails to do so. Regarding CORE, it means that at least two nodes can be chosen in the forwarder set of each packet. \figurename~\ref{figure:throughput8} presents the throughput of BEND, CORE, COPE, non-coding and FlexONC for three lowest BERs in our experiments.

We observe that when BER = $2 \times 10^{-6}$ (i.e., the network condition is almost perfect), most transmitted packets are received by the intended forwarders successfully. Therefore, there hardly exists an opportunity for non-intended forwarders to decode and forward a packet on behalf of the intended forwarder. It is obvious that in such a situation, FlexONC does not show its real power and its throughput is close to BEND. However, as the BER increases, more opportunities for non-intended forwarders are provided and FlexONC's gain over other methods increases significantly.

\begin{figure}[!t]
\centering
\includegraphics[scale=0.45]{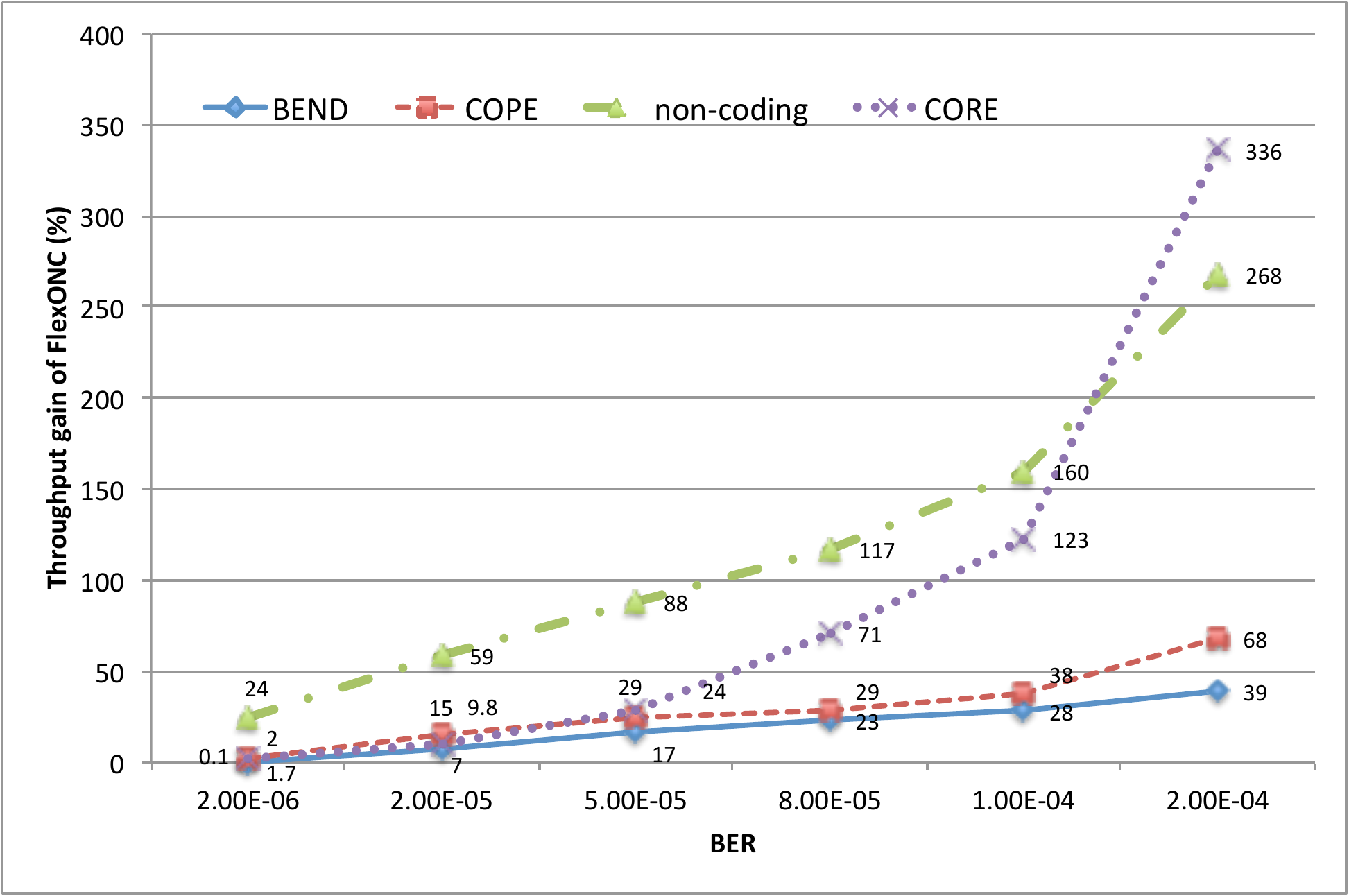} 
\caption{FlexONC's gain over other methods in 8-node topology.}
\label{figure:gain8}
\end{figure}

Furthermore, \figurename~\ref{figure:gain8} presents the performance gain of FlexONC over BEND, CORE, COPE and non-coding for 6 different BER levels, which corroborates our observation. In particular, by increasing the BER, FlexONC becomes more powerful in comparison to the baselines, and its throughput gain increases. The throughput gain of FlexONC over each baseline is calculated as:
\begin{equation}
\label{eq:gain}
\mbox{throughput gain}=\dfrac{Tr(\mbox{FlexONC})-Tr(\mbox{baseline})}{Tr(\mbox{baseline})} \times 100
\end{equation}
where $Tr(x)$ denotes the calculated throughput for scheme $x$. 

As shown in these figures, although at lower BER, CORE's performance is very close to FlexONC's, in lossy networks FlexONC outperforms CORE due to the following reasons. First, in this topology with a small forwarder set, at high BERs many packets are lost without being received by any forwarder. Second, in CORE the packets are broadcasted without any retransmission mechanism to compensate for packet loss.

\subsubsection{Grid Topology}
\begin{figure}
\centering
\begin{subfigure}{.25\textwidth}
  \centering
  \includegraphics[width=0.9\linewidth]{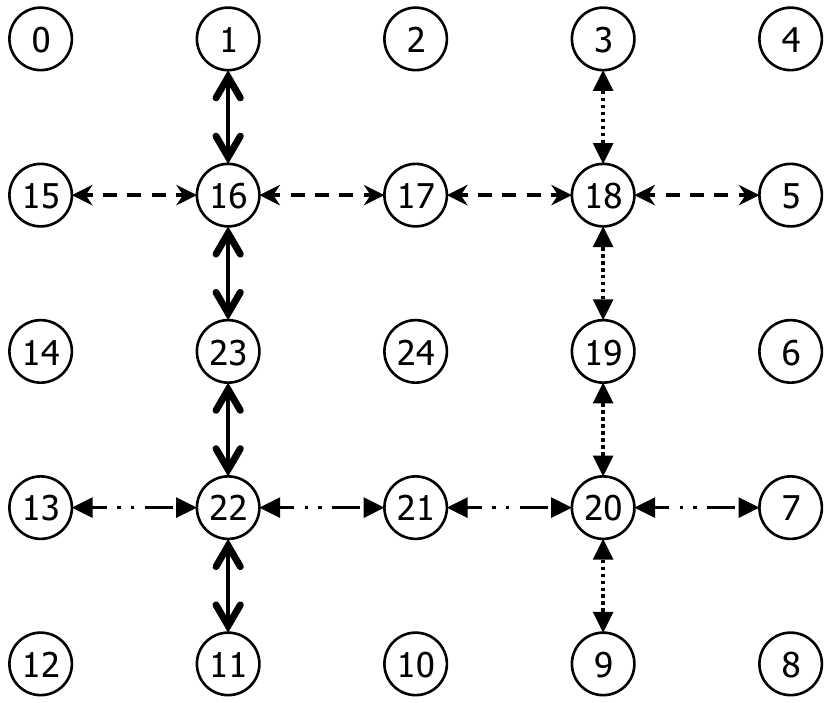}
  \caption{}
  \label{fig:gridTopology1}
\end{subfigure}%
\begin{subfigure}{.25\textwidth}
  \centering
  \includegraphics[width=0.9\linewidth]{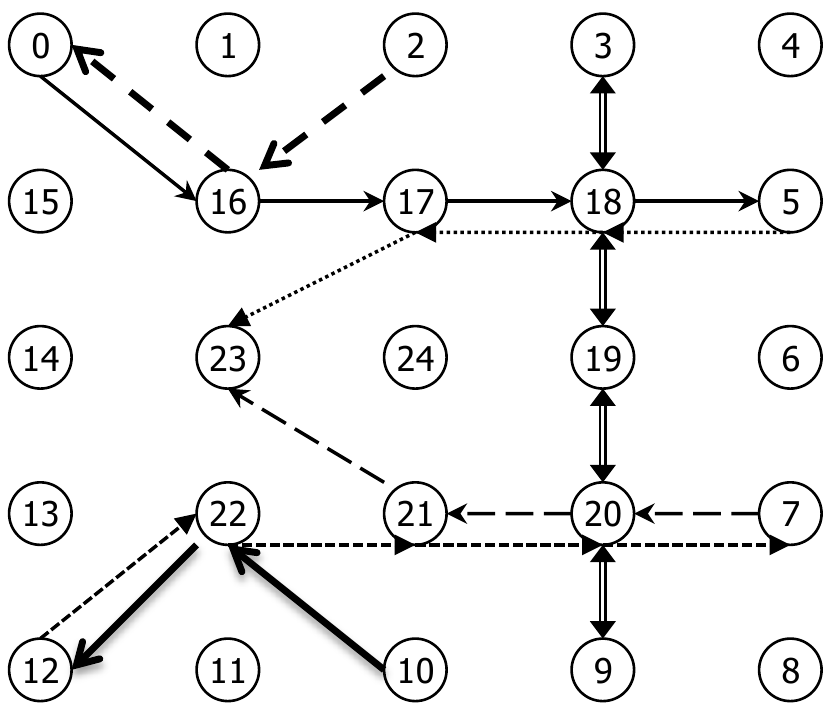}
  \caption{}
  \label{fig:gridTopology2}
\end{subfigure}
\caption{$5 \times 5$ grid topology.}
\label{figure:gridTopology}
\end{figure}

To investigate the performance of FlexONC in a general topology, we test it in a $5 \times 5$ grid, where again the distance between two adjacent nodes is 150 m. 8 different flows with an inter-arrival time of 0.1 s and duration of 150 s transmit packets between Row 2 and Row 4, and also Column 2 and Column 4 of the grid, as shown in \figurename~\ref{fig:gridTopology1}.


\begin{figure}[!t]
\centering
\includegraphics[scale=0.67]{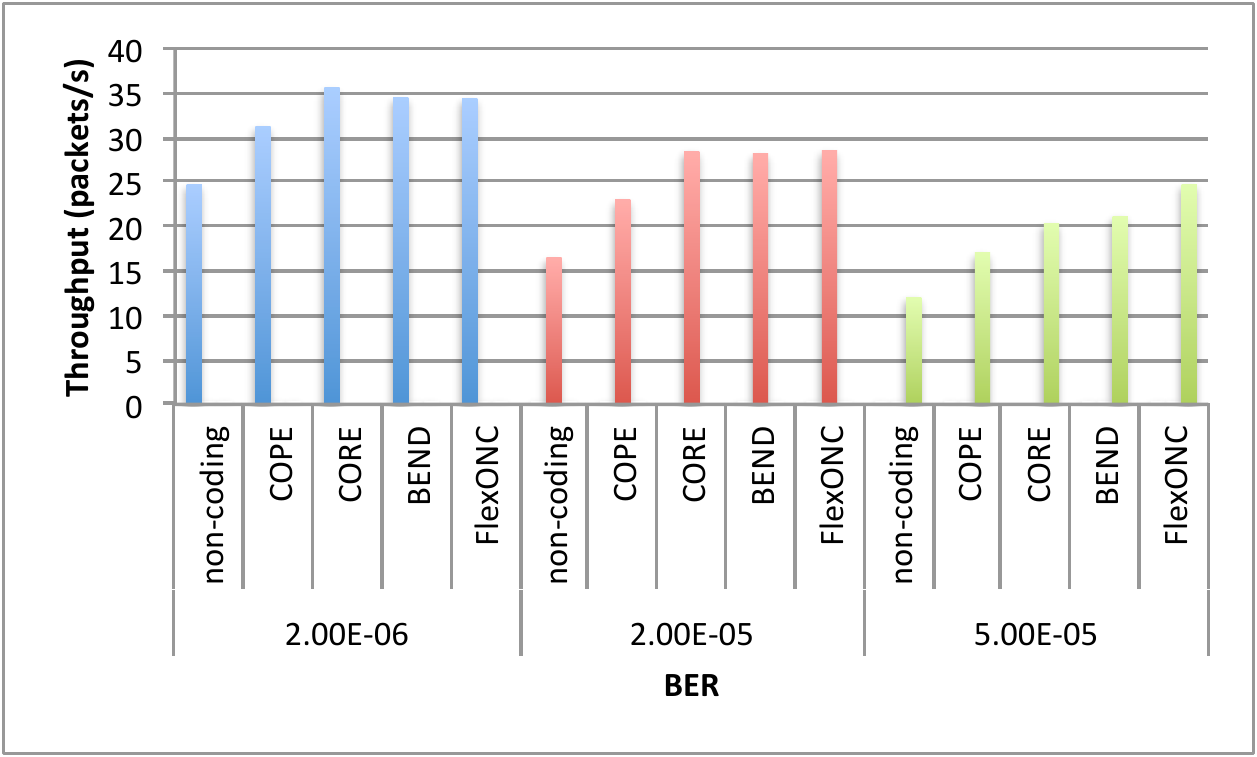}
\caption{Throughput of different methods in the grid topology for different BERs.}
\label{figure:throughputMesh}
\end{figure}

\begin{figure}[!t]
\centering
\includegraphics[scale=0.43]{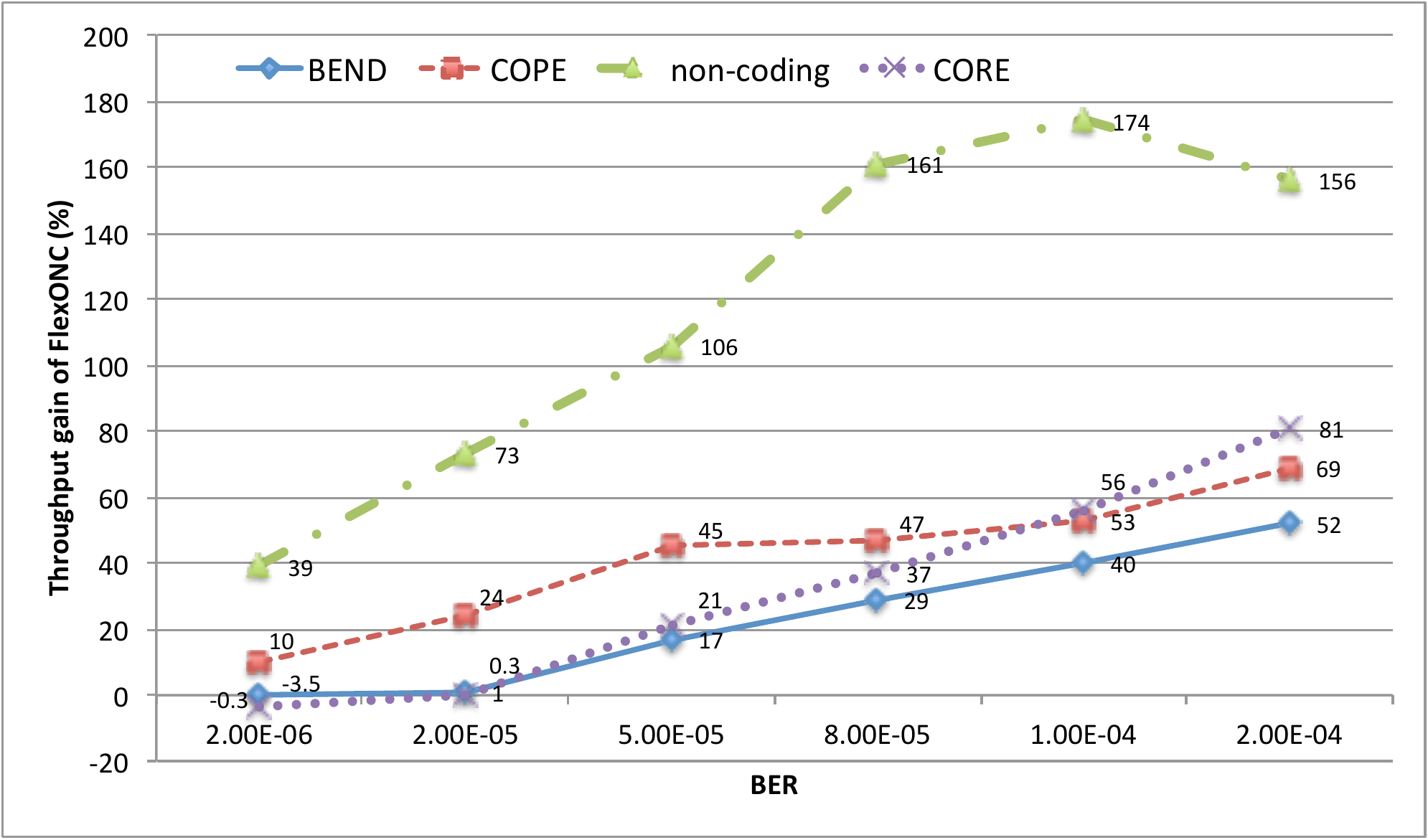} 
\caption{FlexONC's gain over other methods in the grid topology.}
\label{figure:gain25}
\end{figure}

The performance results depicted in \figurename~\ref{figure:throughputMesh} and \figurename~\ref{figure:gain25} again show that at non-trivial BER levels, FlexONC almost always outperforms other methods. In perfect network conditions (BER = $2 \times 10^{-6}$), CORE performs slightly better than FlexONC because there is no intended forwarder in CORE, and it distributes packet transmissions more evenly than FlexONC among possible forwarders. However, as explained earlier, in lossy environments CORE cannot benefit from opportunistic forwarding and network coding as much as FlexONC due to the lack of any retransmission mechanism, especially in such multi-hop routes (i.e., each node should pass at least 4 hops to be delivered to the destination).

In addition, one may notice that by increasing the BER, the throughput gain of FlexONC over CORE increases faster in the 8-node topology in comparison to the grid topology. In fact, the larger forwarder set in the grid topology decreases the probability of packet loss in each transmission.

\subsection{Performance under SwitchRule}
We investigate the effect of \emph{SwitchRule} on the performance of FlexONC in two different scenarios, where at some nodes the common coding conditions may not be sufficient to combine the right packets. First, we compare the throughput of FlexONC in the topology depicted in \figurename~\ref{figure:12node} with different inter-arrival times for cases that the \emph{SwitchRule} functionality is off (i.e., only common coding conditions are used) and is on. We call the latter version of FlexONC, which uses \emph{SwitchRule}, \emph{FlexONC-SR}. In this scenario, 3 flows transmit their packets for 150 s, BER equals $2\times10^{-6}$, and in FlexONC-SR, the NACK threshold to start applying \emph{RecodingRule} is equal to 5.

\begin{figure}[!t]
\centering
\includegraphics[scale=0.4]{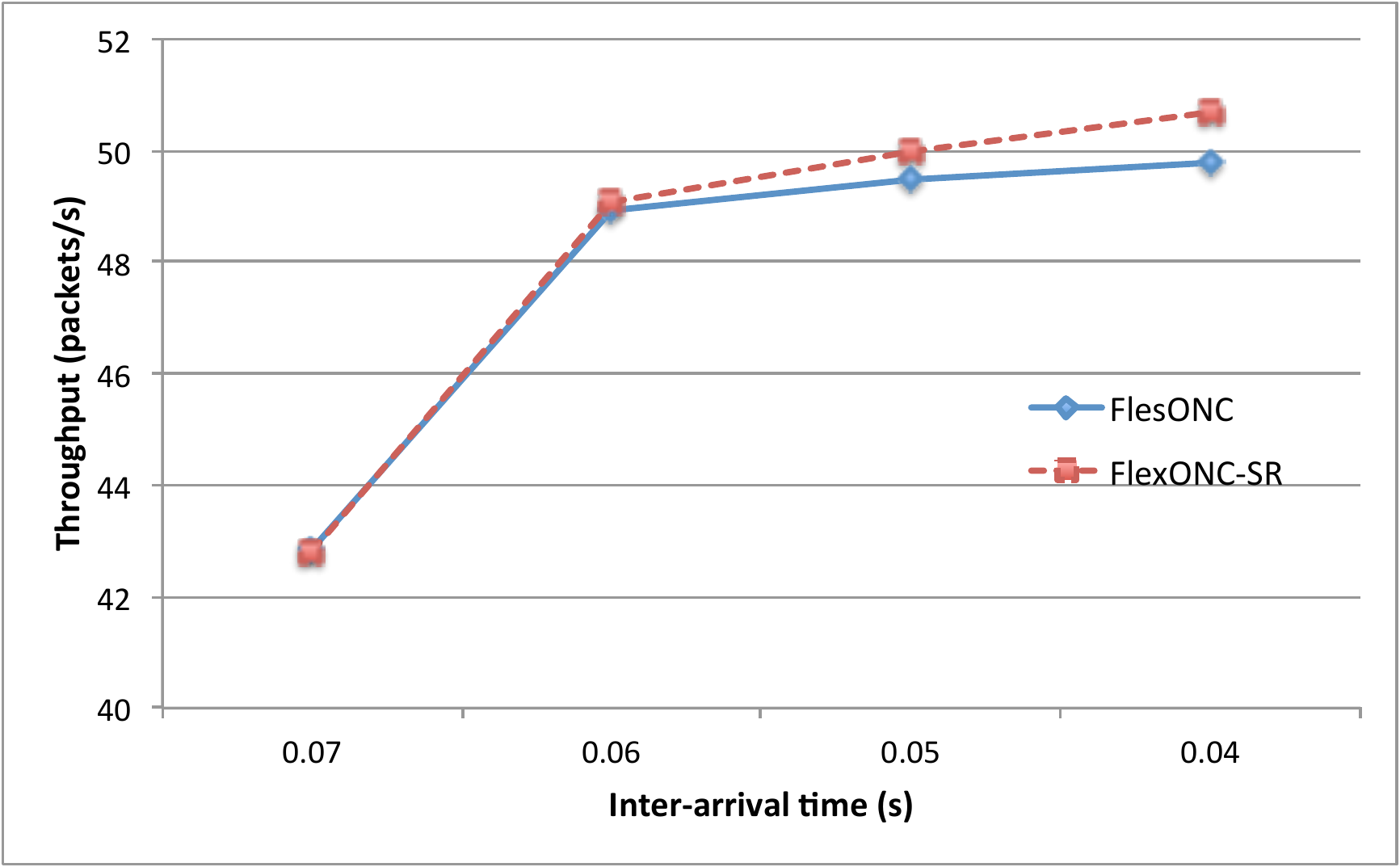} 
\caption{The effect of SwitchRule on the throughput of FlexONC in the topology depicted in \figurename~\ref{figure:12node}.}
\label{figure:FlexONC12}
\end{figure}

\begin{figure}[!t]
\centering
\includegraphics[scale=0.4]{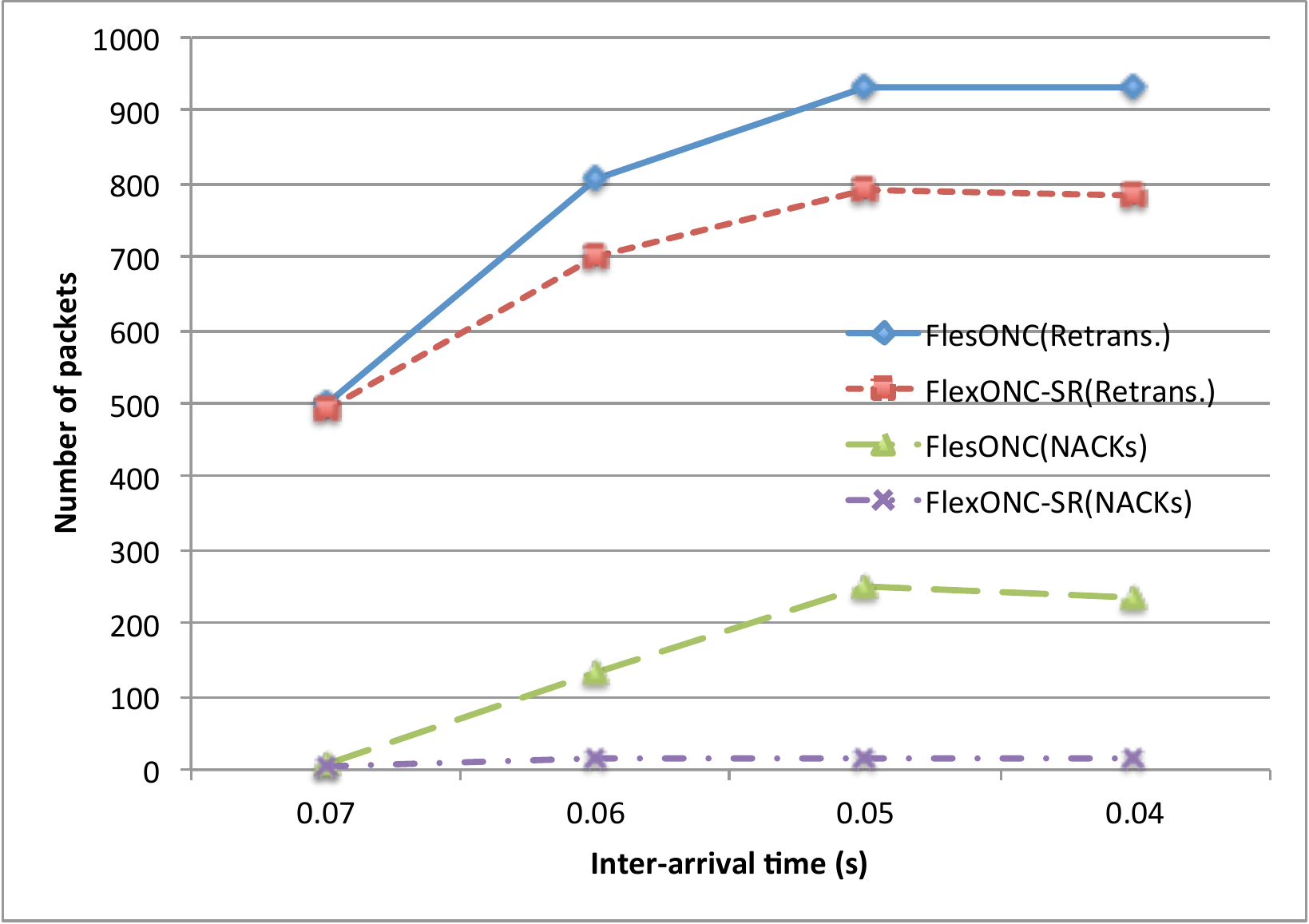} 
\caption{The number of retransmissions and received NACKs with and without applying SwitchRule in FlexONC.}
\label{figure:retrans12}
\end{figure}

As shown in \figurename~\ref{figure:FlexONC12}, although at lower packet arrival rates (i.e., longer inter-arrival time) the performance of FlexONC and FlexONC-SR is close, at higher arrival rates FlexONC-SR can benefit from \emph{SwitchRule} to avoid decoding failures and more retrasnmissions to deliver packets to the destination. As an evidence, \figurename~\ref{figure:retrans12} presents the number of retransmitted packets and the number of received NACKs in both FlexONC and FlexON-SR.  As explained in Subsection~\ref{EXcodingIssue}, the common coding conditions may wrongly decide to combine the decoded-native packets with other packets, and obviously at higher arrival rates, more decoded-native packets are generated (i.e., the probability that the same packet is encoded at different nodes increases).

\begin{figure}[!t]
\centering
\includegraphics[scale=0.45]{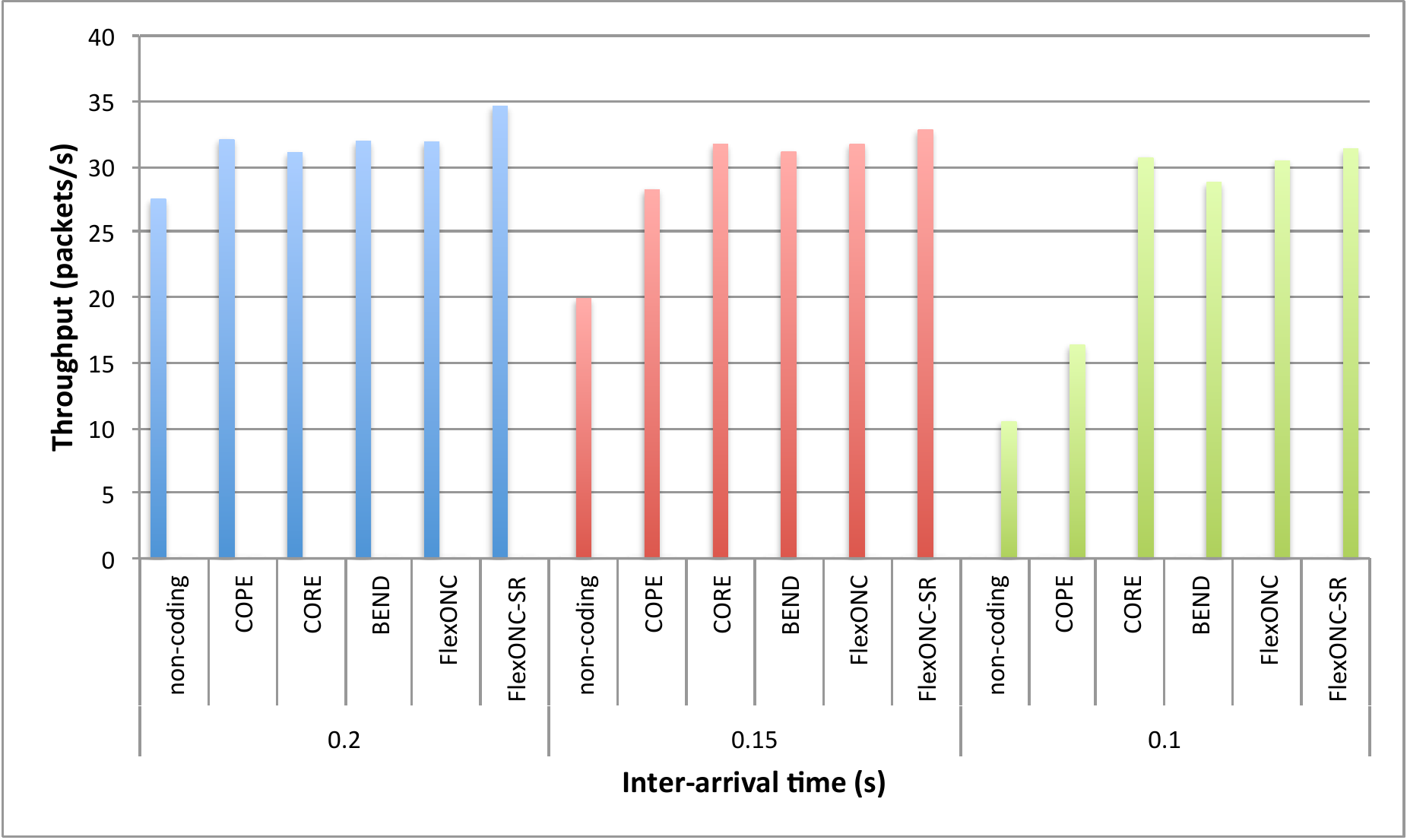} 
\caption{Throughput of different methods in the topology depicted in \figurename~\ref{fig:gridTopology2}.}
\label{figure:throughputSR25}
\end{figure}

We also compare the performance of FlexONC-SR with other baselines in a $5 \times 5$ mesh network with 8 different CBR flows, as depicted in \figurename~\ref{fig:gridTopology2}, with duration of 150 s. As shown in \figurename~\ref{figure:throughputSR25}, although BER is very small ($BER=2 \times 10^{-6}$), FlexONC outperforms other schemes. Moreover, when the functionality of \emph{SwitchRule} is added to FlexONC (i.e., FlexONC-SR), its throughput is even further boosted.


\section{Discussion} \label{section:discussion}
\subsection{Routing Protocol}
In our experiments, we selected DSDV as the routing protocol for its well-known behavior. Moreover, it is a distance-vector approach that makes fewer assumptions about the routing information in comparison to source routing protocols. Therefore, if FlexONC works well with DSDV, it will work with source routing protocols as well. As a matter of fact, choosing DSDV as the routing module does not lose generality of our scheme in a stationary mesh network. We believe choosing any other routing protocol would not make a big difference in FlexONC's performance gain, as long as the routing protocol can be modified in a way that each node contains forwarding information for its neighbors.

\subsection{The End-to-End Delay} \label{delay}
On one hand, FlexONC decreases the delay in forwarding packets and increases the throughput by avoiding packet retransmission when an intended forwarder fails to decode the coded packet, and a non-intended forwarder alternatively passes the packet toward the destination. On the other hand, when more nodes have the responsibility of passing the packet further to the destination, in case of retransmissions, the sender should wait longer for an ACK before it retransmits the packet, and this longer waiting time means longer delay which may lead to a lower throughput. 

Therefore, we face a trade-off here. While the maximum waiting time of the sender is proportional to the number of eligible forwarders, the gain of FlexONC is also related to the number of neighbors of the sender (i.e., more precisely, eligible non-intended forwarders), as well as the probability of intended forwarder's failure in receiving or decoding a coded packet, which is in turn affected by the packet loss probability and BER in the network. The performance result showed that even for a very low BER when the intended forwarder itself can decode and forward the majority of received coded packets and FlexONC does not have much chance to be applied, its performance is comparable to BEND's performance or even better.

\begin{figure}[!t]
\centering
\includegraphics[scale=0.6]{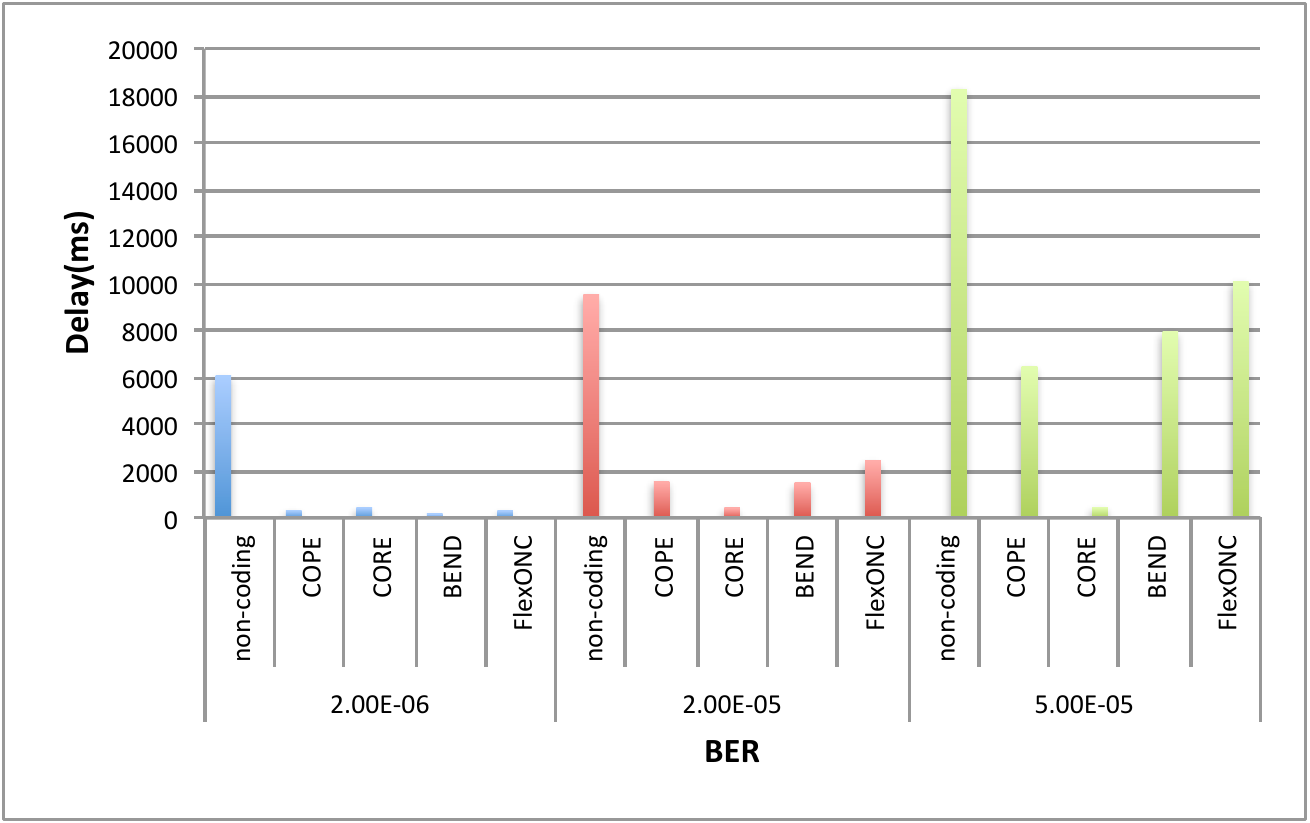}
\caption{The end-to-end delay of different methods in 8-node topology for different BERs.}
\label{figure:delay8}
\end{figure}

\figurename~\ref{figure:delay8} shows the average end-to-end delay of delivered packets in different methods, for the scenario described in Subsection~\ref{subsection:8node}. While the non-coding scheme has the highest average end-to-end delay, the delay in FlexONC is slightly longer than BEND. As explained earlier, the most important reason of this longer delay is that the sender of coded packets in FlexONC waits longer to receive an ACK than in BEND. Therefore, if the packet transmission fails and no ACK is received, BEND's timer, for anticipated ACKs, usually expires earlier than FlexONC's, leading to a faster retransmission in BEND, which can reduce its average end-to-end delay in comparison to FlexONC.

In addition, one may notice that in CORE the end-to-end delay does not vary much over different BERs. While at lower BERs, CORE's delay is longer than that of other coding schemes, at higher BERs its delay is significantly shorter than that of other protocols. The main reason of this shorter and almost constant delay in delivery is the lack of any retransmission mechanism; any packet either is delivered by one transmission or is dropped.

As shown in \figurename~\ref{figure:delay8}, the delay in the non-coding scheme is significantly higher than other methods. The main reason is that coding enables free-riding. In other methods, more than one packet can be combined and sent simultaneously, which means that packets can free-ride on other packets. Therefore, the packets are forwarded faster. In addition, this decreases the queue length at nodes, causing shorter waiting time and consequently shorter delay.

\begin{figure}[!t]
\centering
\includegraphics[scale=0.52]{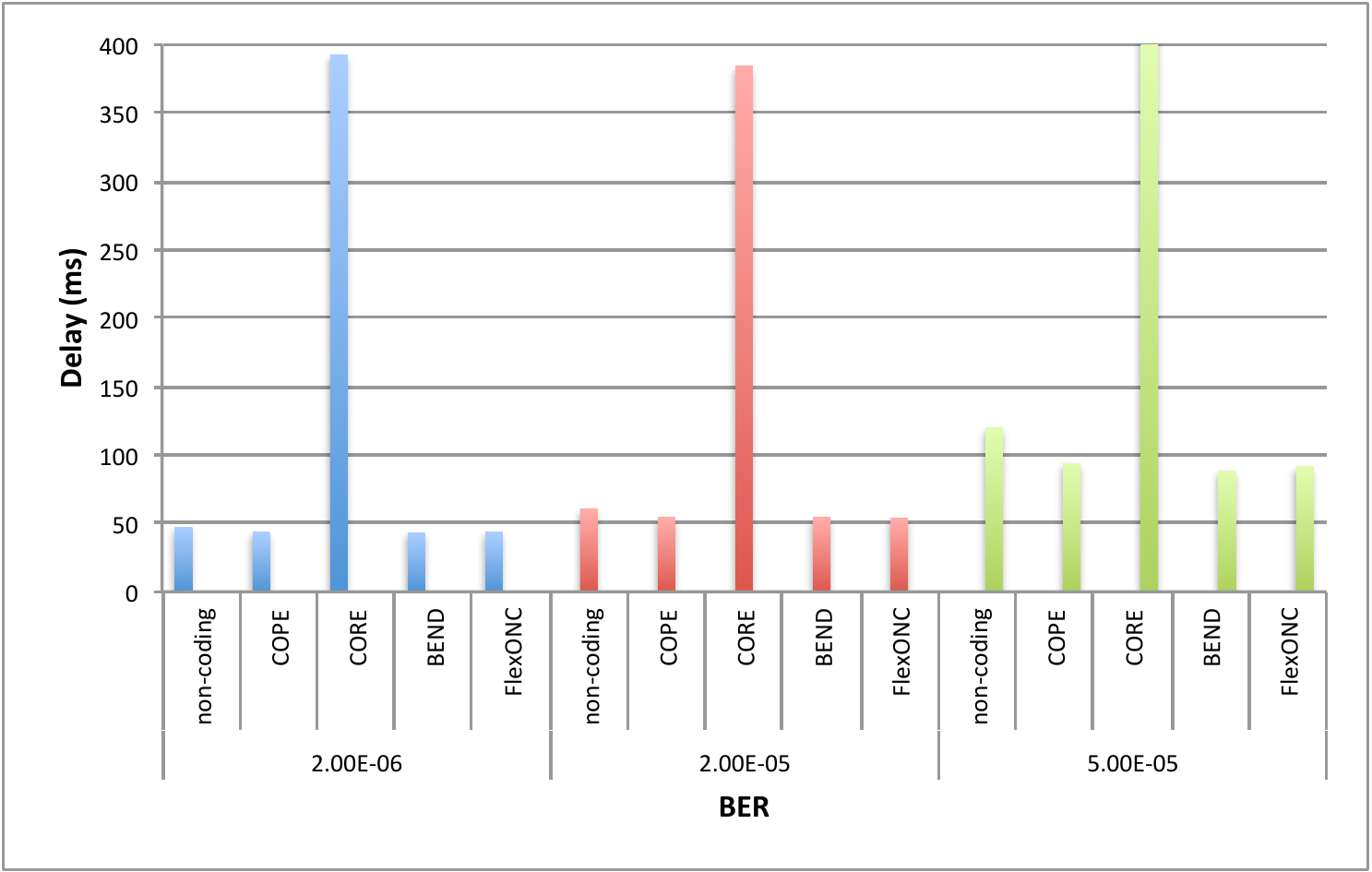}
\caption{The end-to-end delay of different methods in 8-node topology for different BERs with less CBR traffic.}
\label{figure:delaySmaller}
\end{figure}

To verify this explanation we repeat simulations with less CBR traffic with the inter-arrival time of $0.15$ s (instead of $0.07$ s). By increasing the inter-arrival time, less packets are injected to the network per second, which reduces the probability of having more than one packet in the queues, and in turn, creates less coding opportunities at nodes. The results are shown in \figurename~\ref{figure:delaySmaller}, where the delay in non-coding is comparable to the other methods, as the coding schemes provide less free-riding opportunities for the packets.

Furthermore, while this figure justifies the almost constant end-to-end delay in CORE over different BERs, it also shows that the delay in CORE is significantly longer than that of other methods. As mentioned earlier, in this scenario with a small packet arrival rate, the coding opportunities are rare in the network, and most packets are sent natively. To provide higher priority for coded transmissions in CORE, the native packets are delayed before transmission; therefore, forwarding a large number of native packets in this scenario increases the end-to-end delay significantly.

\subsection{Duplicate Packets}

\begin{figure}[!t]
\centering
\includegraphics[scale=0.6]{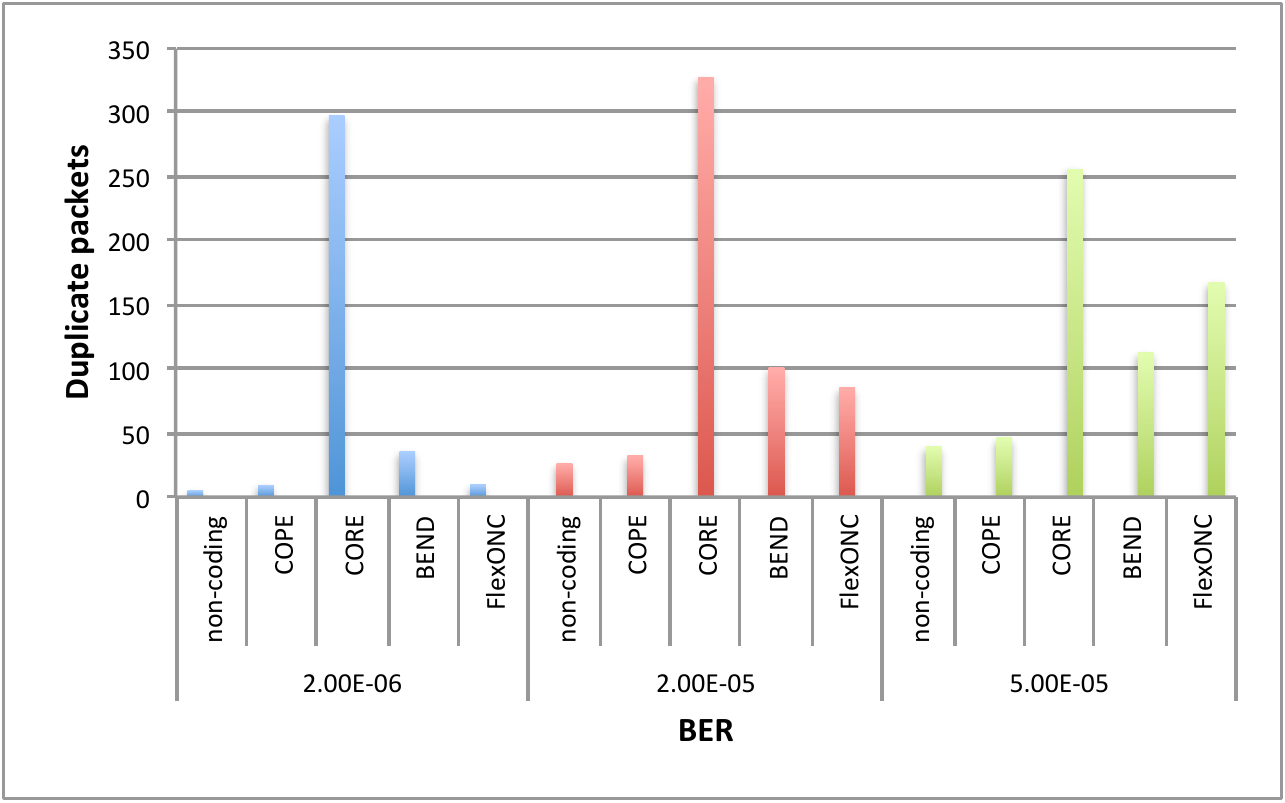}
\caption{Duplicate packets of different methods in 8-node topology for different BERs.}
\label{figure:duplicate}
\end{figure}

As explained in~\cite{BEND-Zhang-CNJournal2010}, since in BEND more nodes cooperate in forwarding packets toward the final destination, it is prone to generating more duplicate packets in case of imperfect collaboration among nodes. The situation in FlexONC could seem even more severe, as it allows non-intended forwarders to cooperate in more ways (i.e., forwarding of not only received native packets, but also received coded packets). To control duplicate packets in FlexONC, we introduced some mechanisms in Subsection~\ref{duplicate}.

\figurename~\ref{figure:duplicate} shows the number of duplicate packets generated by different methods. As shown in this figure, the largest number of duplicate packets are generated at CORE, as nodes should only rely on overhearing other transmissions to avoid duplicate packets. In addition, while the number of duplicate packets in BEND is higher than non-coding and COPE, FlexONC is able to control the number of duplicate packets, especially at lower BERs. The reason could be related to the additional mechanisms introduced in FlexONC to control the number of duplicate packets. However at higher BER=$5\times10^{-5}$, there are more duplicate packets in FlexONC than in BEND because these mechanisms are highly susceptible to the reception of ACKs and at higher BERs the probability of losing ACKs increases.

\subsection{Coding Opportunities}

\begin{figure}[!t]
\centering
\includegraphics[scale=0.47]{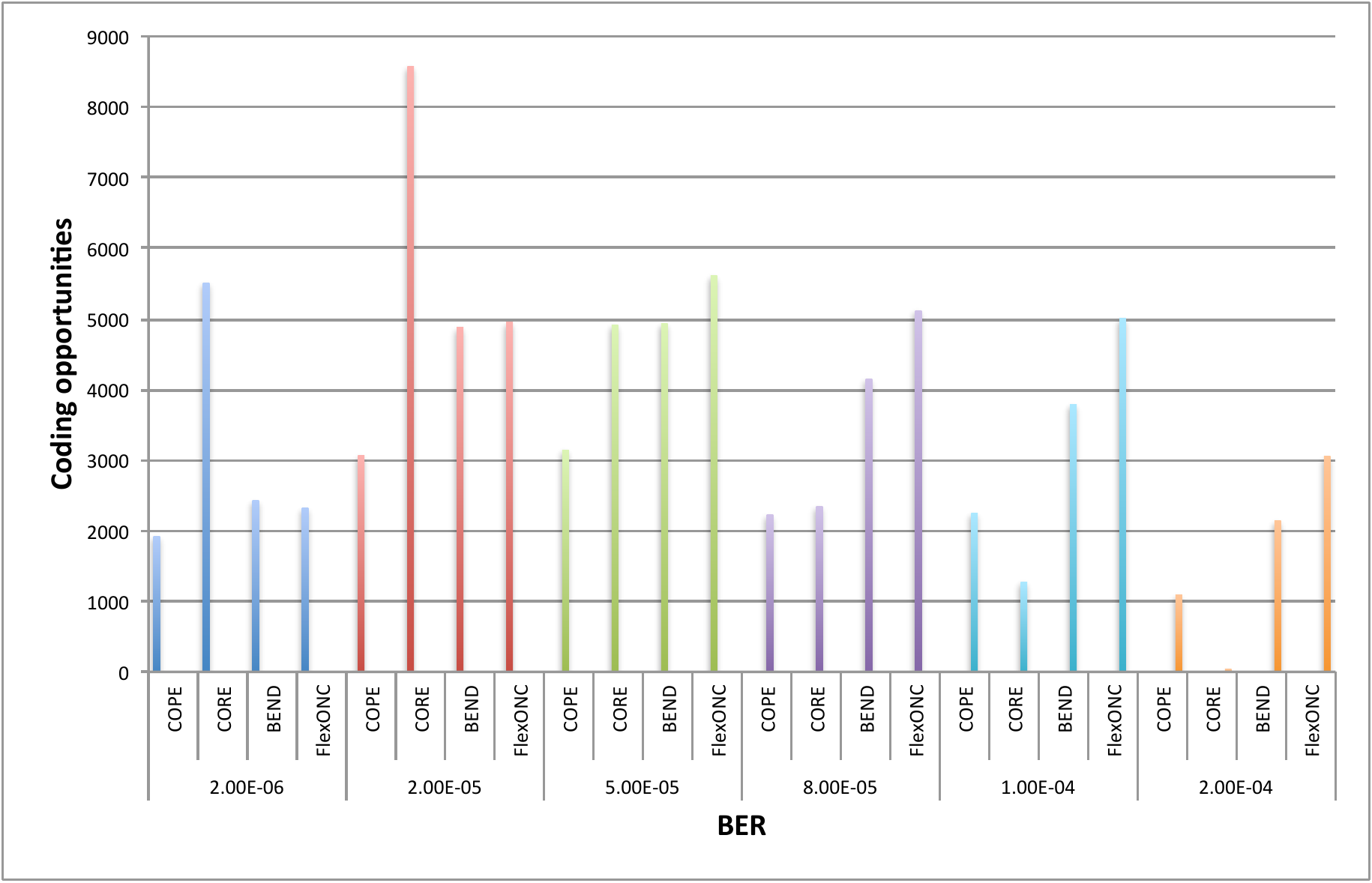}
\caption{Coding opportunities in different methods in 8-node topology for different BERs.}
\label{figure:codingOpp}
\end{figure}

As shown in \figurename~\ref{figure:codingOpp}, at lower BERs the code opportunities at CORE are more than that of FlexONC. However, at higher BERs, FlexONC provides more coding opportunities than other schemes. One may notice that, by increasing BER, first coding opportunities in all methods increases. The reason is that, due to a greater need for retransmission, packets stay longer in the queue and the chance of combining them with the packets of other nodes increases, leading to more coding opportunities. On the other hand, when BER further increases, the number of retransmissions increases significantly; therefore the probability of generating new coding opportunities decreases. That is why for BERs higher than $5\times10^{-5}$ the coding opportunities in the networks drops. In CORE, although there is no retransmission, at higher BERs and in this topology many packets can not go further than one or two hops, which decreases the number of packets in nodes' queues as well as the number of coding opportunities.

\begin{figure}[!t]
\centering
\includegraphics[scale=0.5]{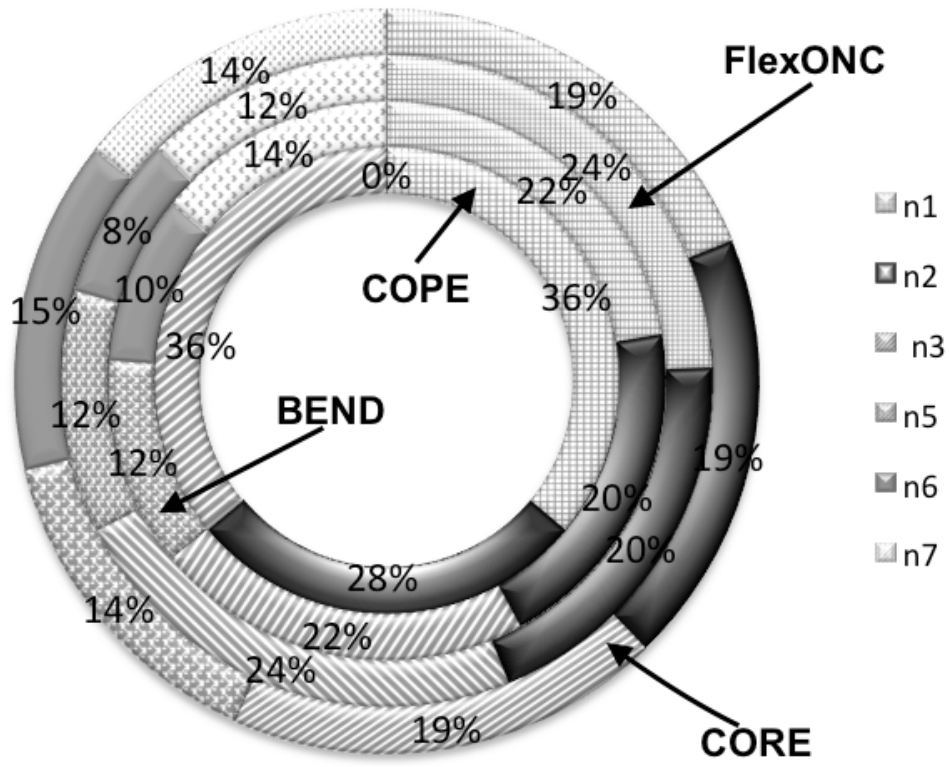}
\caption{The distribution of coding opportunities at different nodes in different methods in 8-node topology.}
\label{figure:codingDist}
\end{figure}

To show the distribution of coding opportunities at different nodes, we run simulations using the topology depicted in \figurename~\ref{figure:8node} and the scenario explained in Subsection~\ref{subsection:8node}, but the route between $N_0$ and $N_4$ is fixed through $N_1$, $N_2$ and $N_3$ for COPE, BEND and FlexONC (i.e., the intended forwarders are $N_1$, $N_2$ and $N_3$). As shown in \figurename~\ref{figure:codingDist}, coding opportunities in COPE are restricted to the intended forwarders; however, other coding schemes use non-intended forwarders (i.e., $N_5$, $N_6$ and $N_7$) to accelerate packet forwarding and provide more coding opportunities. In addition, since in CORE there is no intended forwarder, and possible forwarders are prioritized only based on coding opportunities, the coding opportunities are distributed more evenly in CORE than in other coding schemes.

\subsection{What Happens to Coded Packets in FlexONC?}

\begin{figure}[!t]
\centering
\includegraphics[scale=0.55]{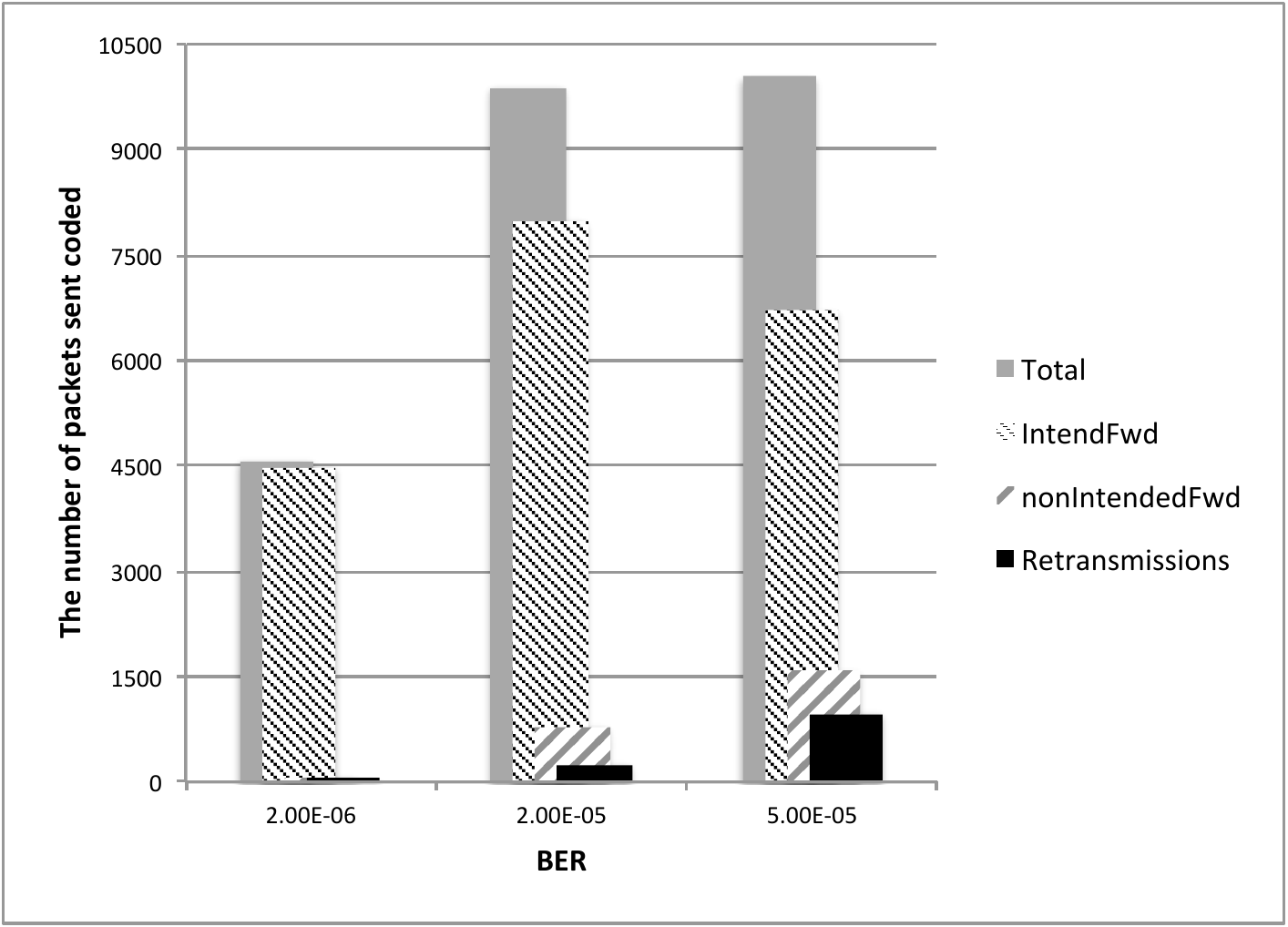}
\caption{What happens to coded packets when BER changes.}
\label{figure:codingInFlexonc}
\end{figure}
To show why by increasing BER FlexONC outperforms other schemes in throughput, we run simulations using the scenario depicted in Subsection~\ref{subsection:8node}, and calculate: 1) the total number of coded packets sent, 2) the number of coded packets received and forwarded by the intended forwarder, 3) the number of coded packets only received and forwarded by one of the non-intended forwarders (i.e., on behalf of the intended forwarder), and 4) the number of coded packets for which the sender does not receive any ACK (or NACK) and retransmits.

As shown in \figurename~\ref{figure:codingInFlexonc}, by increasing BER, intended forwarders receive a smaller percentage of total coded packets sent, and the portion of coded packets which are received only by non-intended forwarders increases. This means that non-intended forwarders can cooperate more effectively in forwarding and be more beneficial. This collaboration among nodes, which increases at higher BER, is the key idea of FlexONC, which leads to increased robustness and higher packet delivery ratio in comparison to the baselines.

\subsection{Packet Delivery Rate} 
Opportunistic forwarding is utilized to increase the probability of successful delivery of a packet as more nodes can help in forwarding packets. In this subsection, we investigate the effect of the number of nodes in the forwarder set, and the link quality on the performance of opportunistic forwarding protocols, especially BEND and FlexONC, for both native and coded packets. We focus on the case with no retransmission first, and the case with retransmission is a natural extension, as we see later. Also, we assume that the nodes in the forwarder set have a perfect coordination mechanism, which means that all nodes in the forwarder set know which one of them forwards the packet. 

Let us denote $p$ as the probability of successful transmission at each link, and $N$ as the average number of nodes in the forwarder set. Then, the probability of successful transmission of a native packet to at least one of the nodes in the forwarder set equals: $p_f^n=1-(1-p)^N$. If a packet traverses $H$ hops in average to be delivered to the destination, in each transmission $N-1$ non-intended forwarders help the intended forwarder except for the transmission to the destination. Then, the probability of successful delivery to the destination can be calculated as: $p_d^n=(1-(1-p)^N)^{H-1}\times p$. It is worth noticing that for $N=1$ (i.e., only one node in the forwarder set of each transmission), $p_d^n=p^H$, which is basically the probability of successful delivery of a packet in traditional forwarding with $H$ hops. Furthermore, when $N$ increases, $p_d^n > p^H$, which shows that by increasing the number of non-intended forwarders (i.e., the nodes in the forwarder set) the packet delivery rate increases.

Regarding coded packets, a received coded packet with $m$ coding partners is decoded successfully if $m-1$ coded partners have already been received. Therefore, the probability of delivery of a coded packet to the next-hop equals $p^m$. As discussed earlier, in BEND coded packets are only forwarded by the intended forwarder (i.e., no opportunistic forwarding). Therefore, the probability of delivery of a coded packet with $m$ coding partners to the destinations in BEND equals $p_d^c(\text{BEND})=(1-(1-p)^N)(p^m)^{H-1}$, given that the source always sends native packets. On the other hand, since FlexONC extends opportunistic forwarding to coded packets as well, the probability of delivery of coded packets to the destination in FlexONC equals: $p_d^c(\text{FlexONC})=(1-(1-p)^N) (1-(1-p^m)^N)^{H-2}p^m$.

To compare the delivery rate in BEND and FlexONC, we focus on the delivery of coded packets, which is different in these two approaches. Assuming that the coding opportunities at both protocols are similar, when the number of non-intended forwarders (i.e., $N$) increases, $p_d^c(\text{FlexONC})$ increases faster than $p_d^c(\text{BEND})$, which shows that the gain obtained by opportunistic forwarding is greater in FlexONC than in BEND. Furthermore, when the link quality is perfect (i.e., $p=1$), the packet delivery ratio for both protocols is the same and independent of $N$, justifying the fact that in perfect network conditions opportunistic forwarding is not beneficial. However, as shown below, in imperfect link qualities (i.e., $p<1$), FlexONC outperforms BEND.
\begin{eqnarray}
  & 0<p<1  \nonumber  \\
  \Rightarrow & 0<p^m<1 \nonumber \\ 
  \xRightarrow{N>1} & (1-p^m)>(1-p^m)^N  \nonumber \\
  \Rightarrow & (1-(1-p^m)) < (1-(1-p^m)^N) \nonumber \\
  \Rightarrow & (p^m)^{H-2}<(1-(1-p^m)^N)^{H-2} \nonumber \\
  \Rightarrow & p_d^c(\text{BEND}) < p_d^c(\text{FlexONC}). \nonumber
\end{eqnarray}

In addition, we can prove in a similar fashion that the performance gap between BEND and FlexONC in terms of the packet delivery rate increases as the link quality decreases. Furthermore, when retransmission is enabled, since $p_d^c(\text{FlexONC}) > p_d^c(\text{BEND})$, each coded packet in FlexONC needs less number of retransmissions to be delivered to the destination, which increases the capacity of the network, and consequently improves the performance.
\subsection{Overall Comparison} \label{subsection:overall}
In this subsection, we provide an overall comparison of FlexONC with other methods, especially BEND, in terms of required storage, packet overhead, computational complexity, delay and throughput. FlexONC provides more coding opportunities, and outperforms other schemes in terms of throughput, especially at higher BERs. Even though having a more powerful protocol may imply increased complexity and overhead, this is not the case of FlexONC, and it is able to keep other metrics such as the end-to-end delay and the number of duplicate packets comparable to other methods, particularly BEND.

Regarding the packet header overhead, while BEND adds the \emph{second-next-hop} field to the packet header of native packets (i.e., four bytes), FlexONC does not need this field. Instead, it adds a bitmap to the header of coded packets to specify eligible forwarders, which is the case in CORE as well. Given the total number of nodes $N$ in the network, the array needs $N$ bits in the packet header, which does not exceed a few bytes in average. Furthermore, to find the forwarder set in each node, CORE adds the geographical-position of the sender and the final destination of each packet to its header, which is not required by FlexONC. 

On the other hand, COPE needs neither the second-next-hop field nor the bitmap since it does not benefit from opportunistic forwarding. Moreover, in FlexONC as well as all other opportunistic forwarding protocols with network coding (e.g., CORE and BEND), all nodes are in promiscuous mode, and store overheard (in addition to intended) packets. Therefore, this overhead is common in all mentioned baselines except for COPE. In fact, in all experiments over different methods, nodes have the same buffer size.

As explained earlier, in FlexONC, in contrast to COPE, CORE and BEND, each node stores the forwarding information of its neighbors. This information is used to control the route followed by packets and avoid them from straying too away from the designated shortest path. If $K$ denotes the maximum number of neighbors of a node in the network, and each entry of the forwarding table needs at most $10$ bytes, the total memory required to store the forwarding information of the neighbors equals $10 \times K\times N$ bytes. Thus, in a network with about $30$ nodes, even if we assume all nodes are connected to each other, the total required storage is less than $9$ KB. On the other hand, while in BEND each node only stores its own forwarding table, the size of this forwarding table is greater than a regular forwarding table, as it stores the IP addresses of the \emph{second-next-hops} in addition to the next-hops themselves. 

All mentioned schemes need to utilize a routing protocol except for CORE as it broadcasts the packets. However, this broadcasting mechanism and lack of retransmission affects the performance of CORE significantly in lossy networks, as shown in the last section. Having routing information of the neighbors in FlexONC only requires adding one extra field to the route advertisement messages of a proactive routing protocol to include the next hop leading to each destination. However, this very small additional routing overhead is not limited to FlexONC; BEND also adds the same field to the route control packets to update \emph{second-next-hop} field in the forwarding table of each node.

Regarding the computational complexity, the most important processes are encoding and decoding which are almost the same in all coding schemes except for CORE. While in FlexONC and other mentioned coding schemes nodes encode the packets in advance immediately after reception, in CORE a packet is encoded when it is going to be transmitted. In addition, to increase the coding gain in lossy environments, CORE introduces a more complicated encoding algorithm in which each node checks all possible coding patterns of the first $K$ packets in its queue. 

In terms of the average end-to-end delay, as explained in Subsection~\ref{delay}, the delay in FlexONC is slightly longer than that in BEND because of the longer maximum waiting time before triggering retransmission of coded packets. Compared to CORE, at lower arrival rates the delay in CORE is significantly longer than that of FlexONC, since CORE delays native transmissions. On the other hand, at higher arrival rates the delay in FlexONC is longer.

Although the experiments in this article are conducted in grid topologies, the benefit of having more \emph{diffusion gain} as well as an additional rule in the coding conditions and having a mechanism to turn it on/off dynamically is still present in general scenarios with random node distribution and flow assignments, and we expect the relative performance among these different methods to be similar to what we have shown here.

\section{Conclusion and Future Work} \label{section:conclusion}
This paper presented FlexONC, an enhancement over BEND, which provides more flexibility and coding opportunities in the network. By utilizing the broadcasting nature of wireless networks, FlexONC is able to spread different flows better than BEND and enable a higher level of cooperation between intended and non-intended forwarders at the link layer in a multi-hop wireless network. Furthermore, by adding an additional rule to the current conditions used to encode the packets in different methods, FlexONC provides more accurate coding conditions, and utilizes \emph{SwitchRule} to apply these coding conditions appropriately and limit decoding failures.

By applying \emph{SwitchRule}, FlexONC is able to adapt coding conditions in different scenarios, and uses a more complete set of rules for encoding when \emph{common coding conditions} are not sufficient. Furthermore, FlexONC benefits from cooperative forwarding especially at higher bit error rates. The performance results show that at higher bit error rates, when an intended forwarder may fail to receive or decode a coded packet and needs its neighbor's help, FlexONC significantly outperforms previous methods like BEND, CORE, COPE and non-coding. Even under an ideal network condition, when intended forwarders usually do not need any help and can decode and forward received coded packets, FlexONC outperforms other schemes because of more precise coding conditions. 


In future work, we plan to provide an analytical model for the combination of opportunistic forwarding and inter-flow network coding in multi-hop wireless mesh networks. Furthermore, in recent years a number of publications have been presented that apply both inter- and intra-flow network coding, but in some limited scenarios~\cite{I2NC-Seferoglu-techReport, CORE-Hansen-CAMAD2013, CORE-Krigslund-VTC2013}. We believe that this combination, if realized carefully, could introduce further improvement in the performance, and represents another way to extend FlexONC.

Moreover, to address the coding condition problem described in this paper, \emph{SwitchRule} is proposed which decides on more precise coding conditions (i.e., \emph{RecodingRule}) in certain scenarios. In future, we plan to propose a scheme that provides nodes with more timely deterministic information and also more accurate probabilistic decisions in encoding. In addition, FlexONC can be extended to include a combination of cooperative forwarding with more powerful detection of coding opportunities beyond a two-hop region~\cite{DCAR-Le-ICDCS2008, FORMpostpone-Guo-VT2011}.

\ifCLASSOPTIONcaptionsoff
  \newpage
\fi



%
\bibliographystyle{IEEEtran} 
\bibliography{citation} 

\begin{thebibliography}{10}
\providecommand{\url}[1]{#1}
\csname url@samestyle\endcsname
\providecommand{\newblock}{\relax}
\providecommand{\bibinfo}[2]{#2}
\providecommand{\BIBentrySTDinterwordspacing}{\spaceskip=0pt\relax}
\providecommand{\BIBentryALTinterwordstretchfactor}{4}
\providecommand{\BIBentryALTinterwordspacing}{\spaceskip=\fontdimen2\font plus
\BIBentryALTinterwordstretchfactor\fontdimen3\font minus
  \fontdimen4\font\relax}
\providecommand{\BIBforeignlanguage}[2]{{%
\expandafter\ifx\csname l@#1\endcsname\relax
\typeout{** WARNING: IEEEtran.bst: No hyphenation pattern has been}%
\typeout{** loaded for the language `#1'. Using the pattern for}%
\typeout{** the default language instead.}%
\else
\language=\csname l@#1\endcsname
\fi
#2}}
\providecommand{\BIBdecl}{\relax}
\BIBdecl

\bibitem{COPE-Katti-IEEEACMTransactions2008}
S.~Katti, H.~Rahul, W.~Hu, D.~Katabi, M.~Medard, and J.~Crowcroft, ``{XORs in
  the Air: Practical Wireless Network Coding},'' \emph{IEEE/ACM Transactions on
  Networking}, vol.~16, no.~3, pp. 497--510, June 2008.

\bibitem{BEND-Zhang-CNJournal2010}
J.~Zhang, Y.~P. Chen, and I.~Marsic, ``{MAC-layer Proactive Mixing for Network
  Coding in Multi-hop Wireless Networks},'' \emph{Computer Networks}, vol.~54,
  no.~2, pp. 196--207, February 2010.

\bibitem{NC-Ahlswede-IEEETransactionsIT2000}
R.~Ahlswede, N.~Cai, S.-Y. Li, and R.~Yeung, ``{Network Information Flow},''
  \emph{IEEE Transactions on Information Theory}, vol.~46, no.~4, pp.
  1204--1216, July 2000.

\bibitem{MORE-Chachulski-SIGCOMM2007}
S.~Chachulski, M.~Jennings, S.~Katti, and D.~Katabi, ``{Trading Structure for
  Randomness in Wireless Opportunistic Routing},'' \emph{SIGCOMM Computer
  Communication Review}, vol.~37, no.~4, pp. 169--180, Aug. 2007.

\bibitem{E-NCP-Lin-INFOCOM2008}
Y.~Lin, B.~Li, and B.~Liang, ``{Efficient Network Coded Data Transmissions in
  Disruption Tolerant Networks},'' in \emph{Proceedings of the IEEE INFOCOM
  2008}, April 2008, pp. 2180--2188.

\bibitem{ICEMAN-Joy-MobiCom2013}
J.~Joy, Y.-T. Yu, M.~Gerla, S.~Wood, J.~Mathewson, and M.-O. Stehr, ``{Network
  Coding for Content-based Intermittently Connected Emergency Networks},'' in
  \emph{Proceedings of the 19th Annual International Conference on Mobile
  Computing and Networking (MobiCom '13)}.\hskip 1em plus 0.5em minus
  0.4em\relax New York, NY, USA: ACM, 2013, pp. 123--126.

\bibitem{NC+DTN-Widmer-SIGCOMM2005}
J.~Widmer and J.-Y. Le~Boudec, ``{Network Coding for Efficient Communication in
  Extreme Networks},'' in \emph{Proceedings of the 2005 ACM SIGCOMM Workshop on
  Delay-tolerant Networking}, ser. WDTN '05.\hskip 1em plus 0.5em minus
  0.4em\relax New York, NY, USA: ACM, 2005, pp. 284--291.

\bibitem{XNCSurvey-Xie-ComNet2015}
\BIBentryALTinterwordspacing
L.~Xie, P.~H. Chong, I.~W. Ho, and Y.~Guan, ``{A Survey of Inter-flow Network
  Coding in Wireless Mesh Networks with Unicast Traffic},'' \emph{Computer
  Networkse}, vol.~91, no.~C, pp. 738--751, Nov. 2015. [Online]. Available:
  \url{http://dx.doi.org/10.1016/j.comnet.2015.08.044}
\BIBentrySTDinterwordspacing

\bibitem{NCSurvey-Iqbal-NCAJournal2011}
M.~A. Iqbal, B.~Dai, B.~Huang, A.~Hassan, and S.~Yu, ``{Review: Survey of
  Network Coding-aware Routing Protocols in Wireless Networks},'' \emph{Journal
  of Network and Computer Applications}, vol.~34, no.~6, pp. 1956--1970, Nov.
  2011.

\bibitem{CORMEN-Islam-COMPUTING2010}
J.~Islam and D.~P.~K. Singh, ``{CORMEN: Coding-Aware Opportunistic Routing in
  Wireless Mesh Network},'' \emph{Journal of Computing}, vol.~2, no.~6, pp.
  71--77, 2010.

\bibitem{CORE-OR-Yan-IEEEWC2010}
Y.~Yan, B.~Zhang, J.~Zheng, and J.~Ma, ``{CORE: a Coding-aware Opportunistic
  Routing Mechanism for Wireless Mesh Networks},'' \emph{IEEE Wireless
  Communications}, vol.~17, no.~3, pp. 96--103, {June} 2010.

\bibitem{CoAOR-Hu-GLOBECOM2013}
Q.~Hu and J.~Zheng, ``{CoAOR: An Efficient Network Coding Aware Opportunistic
  Routing Mechanism for Wireless Mesh Networks},'' in \emph{Proceedings of IEEE
  Global Communications Conference (GLOBECOM'13)}, Dec 2013, pp. 4578--4583.

\bibitem{CAR-Liu-NetSysManage2015}
\BIBentryALTinterwordspacing
H.~Liu, H.~Yang, Y.~Wang, B.~Wang, and Y.~Gu, ``{CAR: Coding-Aware
  Opportunistic Routing for Unicast Traffic in Wireless Mesh Networks},''
  \emph{Journal of Network and Systems Management}, vol.~23, no.~4, pp.
  1104--1124, 2015. [Online]. Available:
  \url{http://dx.doi.org/10.1007/s10922-014-9333-5}
\BIBentrySTDinterwordspacing

\bibitem{FlexONC-Kafaie-ICC2015}
S.~Kafaie, Y.~Chen, M.~H. Ahmed, and O.~A. Dobre, ``{Network Coding with Link
  Layer Cooperation in Wireless Mesh Networks},'' in \emph{IEEE International
  Conference on Communications (ICC'15)}, June 2015, pp. 5282--5287.

\bibitem{InterFlow-Huang-PIMRC11}
L.~Huang and C.~W. Sung, ``{An Iterative Routing Algorithm for Energy
  Minimization in Coded Wireless Networks},'' in \emph{Proceedings of the 22nd
  IEEE International Symposium on Personal Indoor and Mobile Radio
  Communications (PIMRC'11)}, {Sept} 2011, pp. 1124--1128.

\bibitem{NCAQM-Seferoglu-NetCod10}
H.~Seferoglu and A.~Markopoulou, ``{Network Coding-Aware Queue Management for
  Unicast Flows over Coded Wireless Networks},'' in \emph{Proceedings of the
  IEEE International Symposium on Network Coding (NetCod'10)}, June 2010, pp.
  1--6.

\bibitem{BRONC-Guo-CCPR10}
H.~Guo, X.~Liu, Z.~Shi, and X.~Bai, ``{Image Relevance Feedback Retrieval Based
  on Selective Cluster Ensembles},'' in \emph{Proceedings of the Chinese
  Conference on Pattern Recognition (CCPR'10)}, Oct 2010, pp. 1--5.

\bibitem{NCDS-Wang-INFOCOM2013}
S.~Wang, G.~Tan, Y.~Liu, H.~Jiang, and T.~He, ``{Coding Opportunity Aware
  Backbone Metrics for Broadcast in Wireless Networks},'' in \emph{Proceedings
  IEEE INFOCOM}, April 2013, pp. 275--279.

\bibitem{IEEE802.11-2007}
``{IEEE 802.11: Wireless LAN Medium Access Control (MAC) and Physical Layer
  (PHY) Specifications},'' \emph{IEEE Std 802.11-2007 (Revision of Std.
  802.11-1999)}, 2007.

\bibitem{TwoRay-Rappaport}
T.~S. Rappaport, \emph{{Wireless Communications: Principles and Practice}},
  2nd~ed.\hskip 1em plus 0.5em minus 0.4em\relax Prentice Hall, December 2001.

\bibitem{DSDV-Perkins-1994}
\BIBentryALTinterwordspacing
C.~E. Perkins and P.~Bhagwat, ``{Highly Dynamic Destination-Sequenced
  Distance-Vector Routing (DSDV) for Mobile Computers},'' \emph{ACM SIGCOMM
  Computer Communication Review}, vol.~24, no.~4, pp. 234--244, Oct. 1994.
  [Online]. Available: \url{http://doi.acm.org/10.1145/190809.190336}
\BIBentrySTDinterwordspacing

\bibitem{I2NC-Seferoglu-techReport}
H.~Seferoglu, A.~Markopoulou, and K.~Ramakrishnan, ``{Intra- and Inter-Session
  Network Coding for Unicast Flows in Wireless Networks},'' available at
  {arXiv}:1008.5217, Tech. Rep., August 2010.

\bibitem{CORE-Hansen-CAMAD2013}
J.~Hansen, J.~Krigslund, D.~Lucani, and F.~Fitzek, ``{Bridging Inter-flow and
  Intra-flow Network Coding for Video Applications: Testbed Description and
  Performance Evaluation},'' in \emph{Proceedings of the 18th International
  Workshop on Computer Aided Modeling and Design of Communication Links and
  Networks (CAMAD)}, September 2013, pp. 7--12.

\bibitem{CORE-Krigslund-VTC2013}
J.~Krigslund, J.~Hansen, M.~Hundeboll, D.~Lucani, and F.~Fitzek, ``{CORE: COPE
  with MORE in Wireless Meshed Networks},'' in \emph{Proceedings of the IEEE
  Vehicular Technology Conference (VTC Spring)}, {June} 2013, pp. 1--6.

\bibitem{DCAR-Le-ICDCS2008}
J.~Le, J.~Lui, and D.~M. Chiu, ``{DCAR: Distributed Coding-Aware Routing in
  Wireless Networks},'' in \emph{Proceedings of the 28th International
  Conference on Distributed Computing Systems (ICDCS '08)}, June 2008, pp.
  462--469.

\bibitem{FORMpostpone-Guo-VT2011}
B.~Guo, H.~Li, C.~Zhou, and Y.~Cheng, ``{Analysis of General Network Coding
  Conditions and Design of a Free-Ride-Oriented Routing Metric},'' \emph{IEEE
  Transactions on Vehicular Technology}, vol.~60, no.~4, pp. 1714--1727, May
  2011.

\end{thebibliography}

%








\end{document}